\documentclass[
 reprint,
 superscriptaddress,
 amsmath,
 amssymb,
 aps,
 floatfix,
 longbibliography,  %
]{revtex4-2}

\usepackage{bm} %
\usepackage{booktabs} %
\usepackage{graphicx}  %
\usepackage[colorlinks,linkcolor=blue,citecolor=blue,urlcolor=blue]{hyperref} %
\usepackage{siunitx} %
\usepackage{xcolor} %
\usepackage{todonotes} %
\usepackage{xr} %
\setuptodonotes{inline} %

\newcommand{\papertitle}{Phase diagram morphology shapes droplet propulsion in chemical gradients}

\newcommand{\kT}{k_\mathrm{B} T}  %
\newcommand{\vect}{\boldsymbol} %
\newcommand{\diff}{\mathrm{d}}  %

\newcommand{\Eqref}[1]{\mbox{Eq.\hspace{0.25em}\eqref{#1}}}
\newcommand{\Eqsref}[1]{\mbox{Eqs.\hspace{0.25em}\eqref{#1}}}
\newcommand{\figref}[1]{\mbox{Fig.\hspace{0.25em}\ref{#1}}}
\newcommand{\Figref}[1]{\mbox{Fig.\hspace{0.25em}\ref{#1}}}

\newcommand{\secref}[1]{\mbox{section\hspace{0.25em}\ref{#1}}}

\newcommand{\refcite}[1]{\mbox{reference\hspace{0.25em}\cite{#1}}}

\begin{document}

\title{\papertitle}

\author{Stefan Köstler}
\affiliation{Max Planck Institute for Dynamics and Self-Organization, Am Faßberg 17, 37077 Göttingen, Germany}
\affiliation{Institute for Theoretical Physics, University of Göttingen,
Friedrich-Hund-Platz 1, 37077 Göttingen, Germany}

\author{Malcolm Steen}
\affiliation{Max Planck Institute for Dynamics and Self-Organization, Am Faßberg 17, 37077 Göttingen, Germany}
\affiliation{Institute for Theoretical Physics, University of Göttingen,
Friedrich-Hund-Platz 1, 37077 Göttingen, Germany}

\author{David Zwicker}
\email[Contact author: ]{david.zwicker@ds.mpg.de}
\affiliation{Max Planck Institute for Dynamics and Self-Organization, Am Faßberg 17, 37077 Göttingen, Germany}

\date{\today}
\begin{abstract}
Droplets in complex environments often encounter compositional gradients that drive their propulsion. For example, surfactant gradients induce surface-tension-driven propulsion via the Marangoni effect. Here, we show that such propulsion emerges generically, even in the absence of surfactants. Employing a thin-interface approximation, we derive a compact expression for the droplet velocity in terms of droplet size, viscosity, the sensitivity of surface tension to a regulating component, and the chemical potential gradient of that component. Our theory reveals that droplets move toward regions of lower stability, as encoded in the morphology of the phase diagram, particularly near critical points. Numerical simulations confirm these results, which establish a general route to predicting and designing droplet motility from the phase behavior of complex fluids.
\end{abstract}

\maketitle

\section{Introduction}\label{sec:introduction}

Phase-separated droplets are subject to complex environments in biological~\cite{brangwynneGermlineGranulesAre2009,sahaPolarPositioningPhaseSeparated2016,albertiConsiderationsChallengesStudying2019,gouveiaCapillaryForcesGenerated2022,zwickerPhysicsDropletRegulation2025} and synthetic systems~\cite{thutupalliSwarmingBehaviorSimple2011,cejkovaDynamicsChemotacticDroplets2014,herminghausInterfacialMechanismsActive2014,maassSwimmingDroplets2016,lohsePhysicochemicalHydrodynamicsDroplets2020}, where they are exposed to gradients of temperature, material properties, and various chemical species.
Droplets respond to these gradients, either through diffusive mechanisms that translate droplets without momentum transport~\cite{weberDropletRipeningConcentration2017,sahaPolarPositioningPhaseSeparated2016,hafnerReactionDrivenDiffusiophoresisLiquid2024}, or through advective mass flows.
Such mass transport can be driven externally, e.g., via gravity or other body forces, or internally, e.g., by motile active matter~\cite{marchettiHydrodynamicsSoftActive2013} and other forms of activity~\cite{kostlerAdvectionSelectsPattern2026}.
One interesting case involves surface tension gradients~\cite{ratkeInfluenceParticleMotion1985,maassSwimmingDroplets2016,schmittMarangoniFlowDroplet2016}, which transform energy stored in chemical gradients into droplet propulsion.
Understanding this type of propulsion is important for applications in microfluidics, such as lab-on-a-chip devices~\cite{huebnerMicrodropletsSeaApplications2008,baroudDynamicsMicrofluidicDroplets2010}, and for elucidating momentum transport in biomolecular condensates~\cite{brangwynneGermlineGranulesAre2009}.

Droplet propulsion driven by surface tension, also known as Marangoni propulsion, has been studied extensively in the context of surfactants~\cite{maassSwimmingDroplets2016,hardyKineticTheoryCoupled2025,thutupalliSwarmingBehaviorSimple2011,michelinSpontaneousAutophoreticMotion2013,herminghausInterfacialMechanismsActive2014,jinChemotaxisAutochemotaxisSelfpropelling2017,vossChemomechanicalMotilityModes2025}, where surface tension gradients arise from heterogeneous distributions of surfactants that adsorb onto the droplet interface.
Propulsion is also possible without surfactants, e.g., in temperature gradients~\cite{youngMotionBubblesVertical1959,tegzePhaseFieldSimulation2005} or gradients of regulating species~\cite{jambon-puilletPhaseseparatedDropletsSwim2024,furukiMarangoniDropletsDextran2024,dindoChemotacticInteractionsDrive2024}.
Such phenomena have been modeled by assuming that the regulating field either directly controls surface tension~\cite{youngMotionBubblesVertical1959,levichSurfaceTensionDrivenPhenomena1969,michelinSpontaneousAutophoreticMotion2013,schmittMarangoniFlowDroplet2016,vossChemomechanicalMotilityModes2025}, or only affects the capillary stress~\cite{tegzePhaseFieldSimulation2005,dindoChemotacticInteractionsDrive2024,hardyKineticTheoryCoupled2025}.
In these cases, the concentration fields of the regulator and droplet components are not treated on equal footing, and surface tension is prescribed, rather than emerging from interactions between components.
However, recent experiments~\cite{jambon-puilletPhaseseparatedDropletsSwim2024,furukiMarangoniDropletsDextran2024,dindoChemotacticInteractionsDrive2024} have shown that the phase behavior, encoded by interactions between components, strongly informs the motion of droplets in compositional gradients, a feature not captured by current theories.

To understand how interactions between molecular components give rise to gradients in surface tension, and thus droplet propulsion, we develop a framework that connects thermodynamic phase behavior to the interfacial stresses driving advective flows.
We find that a semi-grand-canonical picture is advantageous for obtaining accurate predictions of droplet velocity.
This approach reveals that droplets tend to move toward regions where they would dissolve, a phenomenon known as \emph{dialytaxis}~\cite{jambon-puilletPhaseseparatedDropletsSwim2024}.
We show that the equilibrium phase diagram shapes the direction of droplet motion.

\section{Results}\label{sec:results}

\begin{figure}
    \centering
    \includegraphics[width=0.8\linewidth]{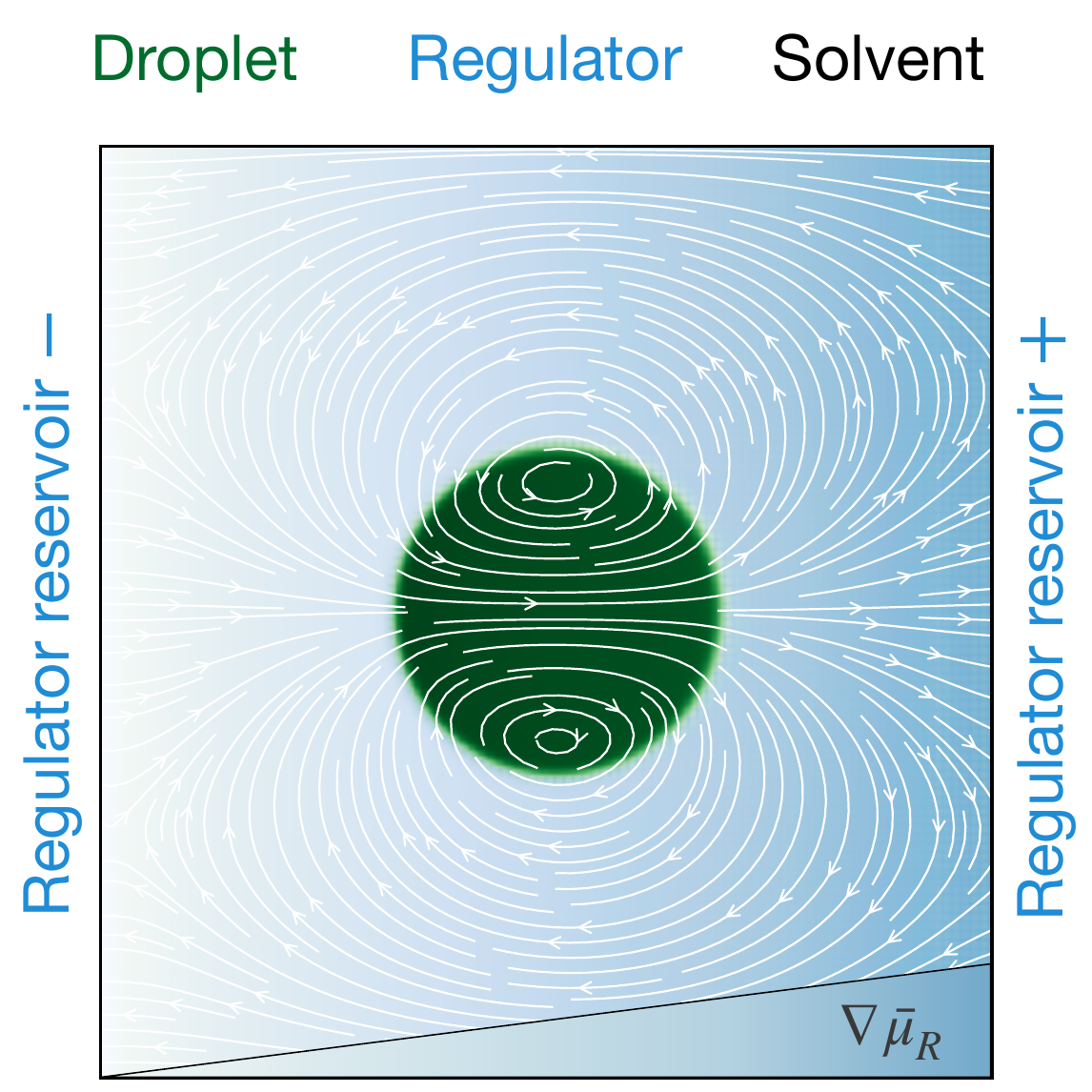}
    \caption{
        \textbf{Droplets move in regulator gradients.} 
        Droplet material (green) segregates from a solvent and a regulator component forming a gradient (blue) due to coupling to different reservoirs on the left ($-$) and right ($+$).
        The resulting interfacial stresses lead to advective flows (white streamlines), which propel the droplet to the right.
    }
    \label{fig:system}
\end{figure}

To study the motion of droplets in compositional gradients, we first introduce numerical simulations of a diffuse-interface model, which we then compare to analytical predictions from a thin-interface model to predict the propulsion velocity.

\subsection{Externally imposed gradients propel droplets}
\label{sec:system_setup}

\subsubsection{Thermodynamics}

Before considering external gradients, we first introduce key concepts in a simple equilibrium system.
We consider an isothermal, incompressible system comprising $N$ different components and a solvent species. The system's state is described by the volume fractions $\phi_1, \ldots, \phi_N$ of all components, whereas the solvent fraction~$\phi_S$ follows from the no-void condition, $\phi_S=1 - \sum_i \phi_i$.
The equilibrium state minimizes the free energy $F=\int f(\{\phi_i\}) \diff V$, where 
\begin{equation}
    f = \frac{\kT}{\nu_0} \Biggl[
        f_0(\{\phi_i\}) 
    + \frac12 \sum_{i,j=1}^N \kappa_{ij} (\partial_\alpha \phi_i) (\partial_\alpha \phi_j)
    \Biggr] %
    \label{eqn:free_energy_density}
\end{equation}
is the free energy density.
Here, $\kT$ is the thermal energy, $\nu_0$ is a reference molecular volume, $f_0$ denotes the local part of the energy density, and the constant, symmetric matrix $\kappa_{ij}$ penalizes gradients.

Minimizing $F$ can lead to multiple coexisting phases~\cite{zwickerPhysicsDropletRegulation2025,qiangScalingLawsPhase2025}.
In thermodynamically large systems, these phases must exhibit chemical equilibrium, which demands equal exchange chemical potentials $\bar{\mu}_j= \nu_j\delta F/\delta \phi_j$, where $\nu_j$ are the molecular volumes of the components.
Moreover, the phases must obey mechanical equilibrium, which implies equal osmotic pressures $\Pi=\sum_i \phi_i\bar\mu_i \nu_i^{-1} - f$.
In finite systems, we additionally need to account for interfaces, across which the pressure $\Pi$ can vary.
Mechanical equilibrium then implies $\partial_\alpha\sigma_{\alpha\beta}=0$, with the associated stress~\cite{zwickerPhysicsDropletRegulation2025}
\begin{equation}
    \sigma_{\alpha\beta} = \Biggl(f - \sum_{i=1}^N \frac{\phi_i \bar{\mu}_i}{\nu_i} \Biggr) \delta_{\alpha\beta} - \sum_{i=1}^N \frac{\partial f}{\partial (\partial_\alpha \phi_i)} \partial_\beta \phi_i \;,
    \label{eqn:stress}
\end{equation}
where the first term accounts for osmotic pressure~$\Pi$, whereas the second term captures interfacial contributions.
In particular, interfaces between phases exhibit a surface tension~$\gamma$, which follows from the excess stress~\cite{kirkwoodStatisticalMechanicalTheory1949}
\begin{equation}
    \gamma = \int_{-\infty}^\infty \bigl[\sigma_{tt}(r) - \sigma_{nn}(r)\bigr] \diff r
    \label{eqn:surface_tension}
    \;,
\end{equation}
where the integral is along the normal direction $n$ across the interface, while $t$ denotes a tangential direction.
Since creating an interface costs an energy $\gamma$ per unit area~\cite{degennesCapillarityWettingPhenomena2004}, equilibrium systems minimize interfacial area, implying spherical droplets.
The pressure inside such a droplet is increased by the Laplace pressure $H\gamma$~\cite{zwickerPhysicsDropletRegulation2025}, where $H=(d-1)/R$ is the sum of principal curvatures of a spherical droplet of radius $R$ in $d$ dimensions.
Consequently, mechanical equilibrium becomes $\Pi_\mathrm{in} = \Pi_\mathrm{out} + H\gamma$.

\subsubsection{Kinetics}
The system is driven out of equilibrium when we apply a gradient by coupling the system to different particle reservoirs at the boundaries at $x=x_-$ and $x=x_+$ (\figref{fig:system}).
We achieve this by imposing exchange chemical potentials,
$\bar\mu_i(x_\pm) = \bar\mu_i^\pm$, describing the exchange of some species~$i$ with solvent across the boundary.
The chemical potential difference $\bar\mu^+_i - \bar\mu^-_i$ then drives a flux of species $i$ through the system.
In contrast, we impose no-flux conditions for all other species, so that their total volume is conserved.

To understand the behavior of this non-equilibrium system, we next investigate the dynamics of all volume fractions~$\phi_i$.
Since all components are locally conserved, $\phi_i$ can only change by center-of-mass advection $v_\alpha$ or by relative fluxes driven by gradients in chemical potentials,
\begin{equation}
    \partial_t \phi_i + v_\alpha \partial_\alpha \phi_i = \partial_\alpha \sum_{j=1}^N \Lambda_{ij} \partial_\alpha \bar{\mu}_j \; ,
    \label{eqn:pde}
\end{equation}
where $\Lambda_{ij}$ are the diffusive mobilities. %
The center-of-mass velocity field $v_\alpha$ satisfies momentum conservation 
\begin{align}
    \partial_\alpha \bigl(\eta e_{\alpha\beta} - \delta_{\alpha\beta} \, p + \sigma_{\alpha\beta} \bigr) = 0 \; , 
    \label{eq:stokes}
\end{align}
where the term in brackets is the total stress comprising viscous stress, given by the product of viscosity~$\eta$ and the rate-of-strain tensor $e_{\alpha\beta}=\partial_\alpha v_\beta + \partial_\beta v_\alpha$, isotropic hydrostatic pressure $p$ to enforce incompressibility $\partial_\alpha v_\alpha=0$, and $\sigma_{\alpha\beta}$ given by \Eqref{eqn:stress}.
In contrast to the equilibrium situation we discussed above, $\partial_\alpha \sigma_{\alpha\beta} = - \sum_i \nu_i^{-1} \phi_i \partial_\beta \bar\mu_i$ is generally non-zero in a system driven out of equilibrium.
Taken together, \Eqsref{eqn:free_energy_density}--\eqref{eq:stokes} can describe droplets in externally imposed gradients.

\subsubsection{Numerical simulations}
\label{sec:numerical_simulations}

To build intuition for the behavior of our system, we consider $N=2$ components and a solvent with equal molecular volumes, where reservoirs impose a gradient of a regulator component $R$, which affects a segregating droplet component $D$.
We describe the interactions between the components by a Flory--Huggins free energy density, $f_0(\phi_D, \phi_R) = \sum_i \phi_i\ln\phi_i + \frac12\sum_{i,j}\chi_{ij}\phi_i\phi_j$, for $i,j=R,D,S$ with solvent fraction $\phi_S = 1-\phi_D-\phi_R$.
We choose the interaction parameters $\chi_{ij}$ such that $D$ segregates from $S$, whereas $R$ does not create its own phase.
To define the free energy density given by \Eqref{eqn:free_energy_density}, we also require the gradient penalty matrix $\kappa_{ij}$ to be positive definite to enable stable interfaces.
For simplicity, we choose  $\kappa_{ij}$ proportionally to the reduced interaction matrix after eliminating the solvent via the no-void condition, $\kappa_{ij} = -\lambda^2(\chi_{ij} - \chi_{iS} - \chi_{Sj})$ for $i,j = D,R$, where $\lambda$ is the interfacial width. %
This choice captures that $\chi_{ij}$ and $\kappa_{ij}$ both originate from the same microscopic interactions; the concrete form can be derived from a lattice model~\cite{cahnFreeEnergyNonuniform1958,naumanNonlinearDiffusionPhase2001}.
Finally, we choose density-dependent mobilities $\Lambda_{ij} = \Lambda_0 (\phi_i \delta_{ij} - \phi_i\phi_j)$, which capture the correct cross-diffusivities in incompressible systems~\cite{zwickerPhysicsDropletRegulation2025,kramerInterdiffusionMarkerMovements1984,maoPhaseBehaviorMorphology2019}.
Here, the bare mobility~$\Lambda_0$ controls the overall diffusive mobilities and allows us to define a fundamental time scale $\tau=\lambda^2/(\Lambda_0 \kT)$, which governs the dynamics of passive droplets.

To impose a regulator gradient, we consider a two-dimensional, rectangular setup with periodic boundary conditions in the $y$-direction, and reservoir boundary conditions for $R$ at the left and right wall, at $x=x_-$ and $x=x_+$ (\figref{fig:system}).
Moreover, we impose vanishing normal derivatives, $\partial_n \phi_i=0$, which implies that all components (including solvent) interact equally with the walls at the reservoirs~\cite{zwickerPhysicsDropletRegulation2025}.
We solve \Eqref{eqn:pde} using a finite difference scheme, whereas the velocity field $v_\alpha$ is determined using a projection method; see Appendix \secref{sec:SInumerics}.
As expected, we find that the regulator exhibits a gradient between the reservoirs, whereas the droplet component is strongly enriched in a near-spherical droplet (\figref{fig:system}).
The associated velocity field~$v_\alpha$ exhibits two vortices, which propel the droplet toward the right, i.e., toward larger regulator concentrations.
It has been proposed that this tendency is caused by the fact that the regulator tends to dissolve the droplet~\cite{jambon-puilletPhaseseparatedDropletsSwim2024}.
Using a detailed analysis of droplets in gradients, we will show  that this is indeed generally the case.

\subsection{Surface tension gradients set droplet velocity}

\begin{figure}
    \centering
    \includegraphics[width=\linewidth]{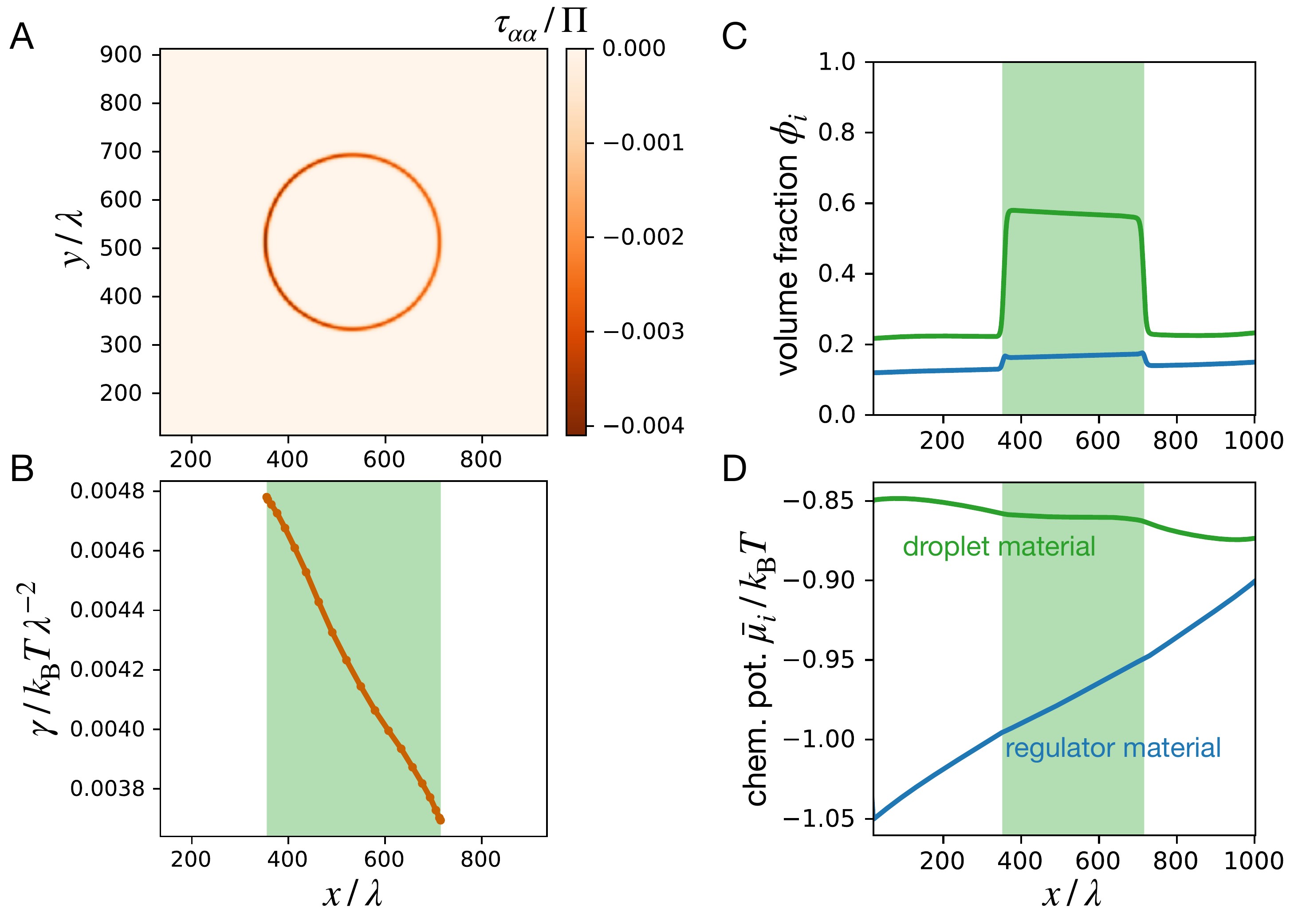}
    \caption{
        \textbf{Regulator gradient controls surface tension.}
        (A)~Simulation snapshot of the trace of capillary stress, $\tau_{\alpha\alpha} = -\frac{\kT}{\nu_0}\sum_{ij} \kappa_{ij}(\partial_\alpha \phi_i)(\partial_\alpha \phi_j)$, compared to osmotic pressure~$\Pi$.
        (B) Surface tension $\gamma$ determined via \Eqref{eqn:surface_tension} as a function of distance~$x$ along the gradient.
        \mbox{(C, D)} Volume fractions $\phi_i$ (panel C) and chemical potentials $\bar\mu_i$ (panel D) as a function of $x$ for droplet material ($i$=$D$, green) and regulator ($i$=$R$, blue).
        Shown are cuts through the center of the droplet; $\bar{\mu}_D$ has been shifted by $-0.8 \, \kT$ for visibility.
        \mbox{(A--D)} Green shaded region marks droplet. Model parameters are $t = 4 \cdot 10^4 \, \tau$, $R_0 = 180 \lambda$, $\Delta \bar{\mu}_R = 0.15 \, \kT$, $\chi_{DS} = 2.5$, $\chi_{DR} = 1.3$, $\chi_{RS} = 1.7$, $\eta = 0.2 \, / \, (\lambda \, \Lambda_0)$,
        $\tau = \lambda^2/(\Lambda_0 \, \kT)$, $L_x = 984 \, \lambda$, and $L_y = 1024 \, \lambda$.
    }
    \label{fig:profiles}
\end{figure}

To understand the propulsion of the droplet, we  analyze the velocity field $v_\alpha$, which is driven by gradients in the stress tensor~$\sigma_{\alpha\beta}$ given by \Eqref{eqn:stress}.
Since contributions from the osmotic pressure~$\Pi$ can be absorbed by the hydrostatic pressure $p$, we write $\sigma_{\alpha\beta} = -\Pi\delta_{\alpha\beta} + \tau_{\alpha\beta}$, to emphasize that the velocity field is determined by
the gradient contribution $\tau_{\alpha\beta} = - \frac{\kT}{\nu_0}\sum_{i,j} \kappa_{ij}(\partial_\alpha \phi_i)(\partial_\beta \phi_j)$.
Since compositional gradients in the bulk are small compared to interfacial gradients, $\tau_{\alpha\beta}$ is localized to the droplet interface (\figref{fig:profiles}A).
Consequently, $v_\alpha$ is driven by interfacial effects, and we hypothesize that surface tension gradients drive the flow via the Marangoni effect.

To show that $v_\alpha$ is driven by surface tension gradients, we introduce a thin-interface approximation, exploiting the localized stress at the interface.
Far away from the interface, the velocity field obeys
\begin{align}
    \partial_\alpha\bigl(\eta e_{\alpha\beta} - P\delta_{\alpha\beta}\bigr) &= 0
    \label{eqn:stokes_homogeneous}
    \;,
\end{align}
where $P = p +  \Pi$ absorbs the osmotic pressure~$\Pi$ and enforces incompressibility $\partial_\alpha v_\alpha=0$.
To obtain a velocity field~$v_\alpha$ everywhere, \Eqref{eqn:stokes_homogeneous} is solved separately inside and outside the droplet with appropriate boundary conditions.
For simplicity, we assume that the pressure $P$ is uniform far away, implying that flows are exclusively driven by the stress at the interface and not by externally applied pressure gradients.
At the droplet interface, we assume continuous velocity fields ($v_\alpha^\mathrm{in} = v_\alpha^\mathrm{out}$) and that the interface is in local equilibrium.
This allows us to identify jump conditions for the viscous stress (Appendix \secref{sec:SIstressinterface} and \refcite{andersonDIFFUSEINTERFACEMETHODSFLUID1998}),
\begin{subequations}
\label{eqn:stokes_bcs}
\begin{align}
    \bigl[\eta e_{nn} - P\bigr]_\mathrm{in}^\mathrm{out} &= H \gamma
\\
    \bigl[\eta e_{nt}\bigr]_\mathrm{in}^\mathrm{out} &= -\nabla_t \gamma
    \;,
\end{align}
\end{subequations}
while $e_{tt}$ follows from incompressibility.
The first condition states that the jump of the normal component of the viscous stress across the interface is given by the Laplace pressure~$H\gamma$, which is proportional to the sum of principal curvatures of the interface, $H$.
In contrast, tangential stresses follow from the surface gradient $\nabla_t$ of the surface tension~$\gamma$. %
Note that the boundary conditions given by \Eqref{eqn:stokes_bcs} can also be interpreted as a force density field~$f_\beta$ localized at the interface, so that $v_\alpha$ is governed by $\partial_\alpha\bigl(\eta e_{\alpha\beta} - P\delta_{\alpha\beta}\bigr) = f_\beta$ in the entire system.
Taken together, in the thin-interface approximation, the velocity field~$v_\alpha$ is entirely controlled by the surface tension along the interface.

To determine the droplet velocity~$v^\mathrm{drop}_\alpha$, we define the droplet as an isocontour of the associated volume fraction field $\phi_D$.
The velocity is then given as the time derivative of the center-of-mass of the contour.
Assuming that the contour is predominantly moved by advection, we find
$v^\mathrm{drop}_\alpha = V_\mathrm{drop}^{-1} \oint \, r_\alpha v_n \diff A$, where the integral is over the surface of the droplet of volume $V_\mathrm{drop}$~\cite{kulkarniEffectiveSimulationsInteracting2023}.
In the case of a near-spherical droplet in a system of constant viscosity~$\eta$, this simplifies to~\cite{schmittMarangoniFlowDroplet2016,yoshinagaSpontaneousMotionDeformation2014}
\begin{align}
    v^\mathrm{drop}_\alpha &= - \frac{a_d}{\pi \eta} \int \diff\Omega \ \hat{r}_\alpha \gamma
    \label{eqn:drop_speed}
    \;,
\end{align}
where $\hat{r}_\alpha$ is the radial unit vector in spherical coordinates centered at $x^\mathrm{drop}_\alpha$ and the integral is taken over the whole solid angle~$\Omega$.
Here, $a_2 = \frac{1}{8}$ and $a_3 = \frac{1}{10}$ capture pre-factors based on dimension $d$.
If the viscosity $\eta_\mathrm{in}$ inside the droplet differs from the viscosity~$\eta_\mathrm{out}$ outside, the entire prefactor in front of the integral differs, e.g., $[-2\pi (2\eta_\mathrm{out} + 3\eta_\mathrm{in})]^{-1}$ for $d=3$~\cite{schmittMarangoniFlowDroplet2016}.
Note that $v^\mathrm{drop}_\alpha$ vanishes for a solid sphere ($\eta_\mathrm{in} \to \infty$), where instead diffusiophoresis becomes relevant~\cite{marbachOsmosisMolecularInsights2019,andersonColloidTransportInterfacial}.
In contrast, for a fluid-like droplet, \Eqref{eqn:drop_speed} shows that variations in surface tension~$\gamma$ set the droplet velocity.

\subsection{Local chemical potential sets surface tension}
\label{sec:surface_tension}

To use \Eqref{eqn:drop_speed} to determine the droplet velocity $v^\mathrm{drop}_\alpha$, we must determine how surface tension~$\gamma$ varies along the droplet interface.
Numerical simulations indicate that $\gamma$ depends roughly linearly with the distance along the gradient (\figref{fig:profiles}B).
Since $\gamma$ is determined via \Eqref{eqn:surface_tension}, the variation of $\gamma$ must originate from variations of the thermodynamic variables.
In principle, we could thus express $\gamma$ as a function of volume fractions $\phi_i$, the chemical potentials $\bar\mu_i$, or a combination since all fields vary in space (\figref{fig:profiles}C,D).
To identify a natural description, we consider global conservation:
The overall amount of droplet material is conserved, suggesting a canonical ensemble with $\phi_D$ as the natural variable.
In contrast, the regulator is controlled by external reservoirs, corresponding to a grand-canonical ensemble and natural variable $\bar\mu_R$.
We thus hypothesize that $\bar\mu_R$ controls the coexisting fractions~$\phi_D$ of droplet material, and thus the associated surface tension~$\gamma$, at each point in space.

\begin{figure}
    \centering
    \includegraphics[width=\linewidth]{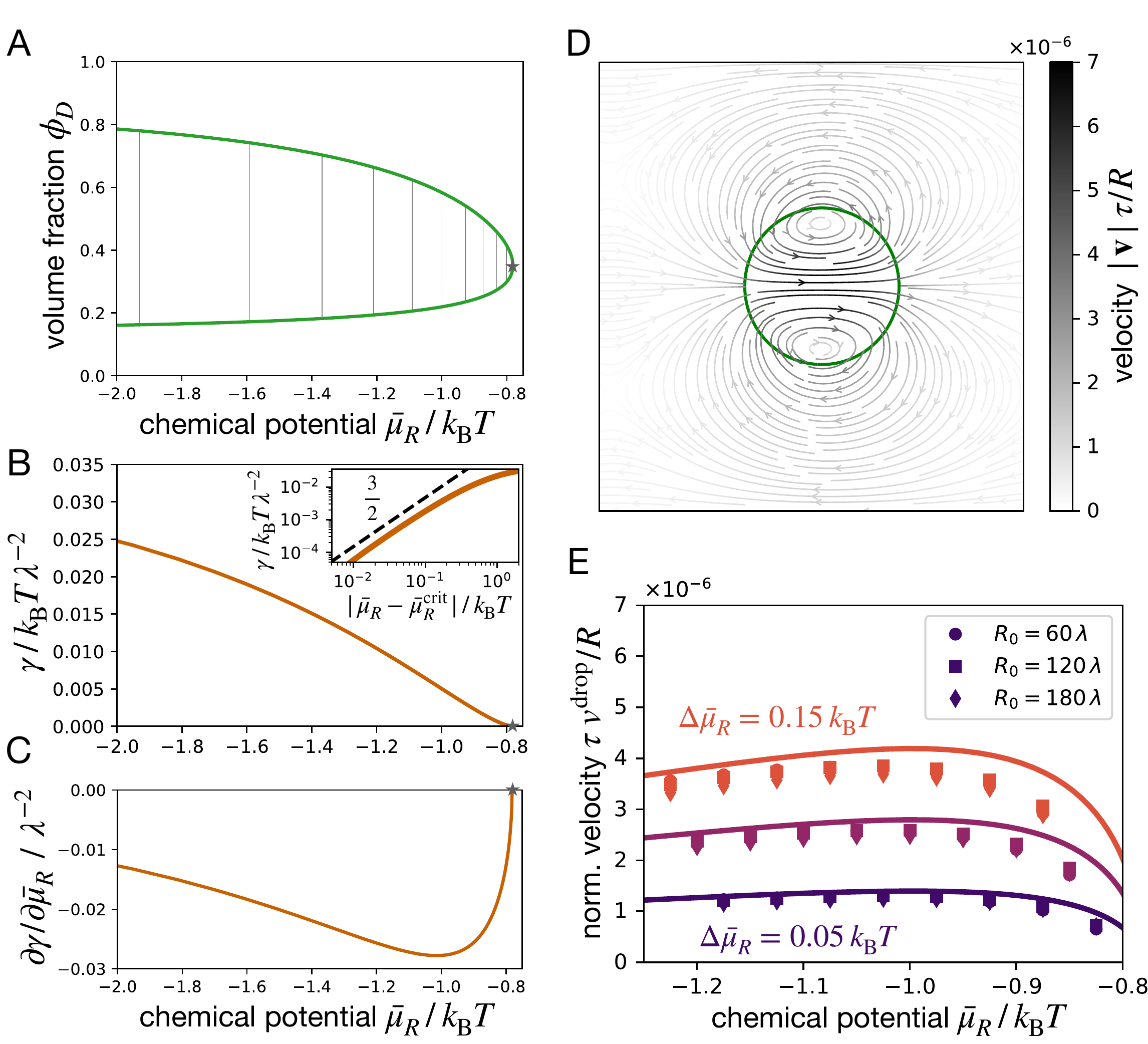}
    \caption{
        \textbf{Local chemical potential sets surface tension.}
        (A) Semi-grand-canonical phase diagram as a function of the fraction $\phi_D$ of droplet material and the chemical potential $\bar{\mu}_R$ of regulator.
        Tie lines (gray) connect coexisting fractions of droplet material (green) for $\bar\mu_R$ smaller than the critical value (star)
        (B)
        Surface tension $\gamma$ as a function of $\bar{\mu}_R$. 
        Inset indicates critical scaling $\gamma \sim |\bar{\mu}_R - \bar{\mu}_R^\mathrm{crit}|^{3/2}$.
        (C) Derivative $\partial \gamma/\partial \bar\mu_R$ as a function of $\bar\mu_R$.
        (D) Streamlines and magnitude (gray scale) of velocity field $v_\alpha$ corresponding to simulation shown in \figref{fig:profiles}.
        Green line indicates droplet interface defined as $\phi_D = 0.42$
        (E) Normalized speed $v^\mathrm{drop}/R$ as a function of $\bar{\mu}_R$ at droplet center for different initial droplet radii $R_0$ and chemical potential differences: $\Delta \bar{\mu}_R = 0.05 \, \kT$ (dark blue), $\Delta \bar{\mu}_R = 0.1 \, \kT$ (purple) and $\Delta \bar{\mu}_R = 0.15 \, \kT$ (orange). 
        Lines show theoretical predictions given by \Eqref{eq:dropletVelocity} for $\partial_\alpha\bar\mu_R$ estimated according to Appendix \secref{sec:estimation}.
        (A--E)
        Additional model parameters are given in \figref{fig:profiles}.
    }
    \label{fig:surfacetension}
\end{figure}

To see whether we can determine surface tensions by focusing on the chemical potential~$\bar\mu_R$ of the regulator, we next determine the associated phase diagram.
\figref{fig:surfacetension}A shows tie lines (thin gray lines) connecting coexisting compositions of droplet material (green line) as a function of $\bar\mu_R$.
In this semi-grand-canonical case, tie lines must be vertical to have identical $\bar\mu_R$ in both phases.
This also implies that $\bar\mu_R$ selects a unique pair of coexisting compositions, which would not have been the case in a canonical phase diagram where we vary $\phi_R$.
In our example, raising $\bar\mu_R$ lowers the compositional difference until phase separation ceases completely at the critical point.
This influence of $\bar\mu_R$ is directly visible in our spatially resolved simulations (\figref{fig:profiles}C).
Since $\bar\mu_R$ uniquely determines coexisting phases, it also governs the full interface profiles if $\kappa_{ij}$ is given.
We can thus use simple thermodynamic equilibrium calculations (see Appendix \secref{sec:SIeqCalc}), and \Eqref{eqn:surface_tension} to determine the surface tension~$\gamma$ as a function of $\bar\mu_R$ (\figref{fig:surfacetension}B).
Note that such an association would be impossible if we focused on the regulator fraction~$\phi_R$ in a canonical ensemble since $\phi_R$ changes strongly across the interface.
Taken together, the semi-grand-canonical approach allows us to associate coexisting fractions~$\phi_D$ and surface tension with each value of $\bar\mu_R$ and thus each position $x_\alpha$.

Using the dependence of surface tension~$\gamma$ on the regulator potential $\bar\mu_R$, we can further simplify the expression for the droplet velocity $v^\mathrm{drop}_\alpha$.
In particular, assuming that the regulator gradient $\bar\mu_R(x_\alpha)$ is quasi-stationary and that variations in $\gamma(\bar\mu_R)$ and $\bar\mu_R(x_\alpha)$ are small, we have
    $\gamma(x_\alpha) \approx \gamma\bigl(\bar\mu_R(x^\mathrm{drop}_\alpha)\bigr) + \frac{\partial \gamma}{\partial \bar{\mu}_R} (\partial_\alpha \bar{\mu}_R)  r_\alpha $,
with $r_\alpha = x_\alpha - x^\mathrm{drop}_\alpha$ for $|r_\alpha| = R$.
Using \Eqref{eqn:drop_speed}, we thus find for two dimensions
\begin{equation}
    v^\mathrm{drop}_\alpha \approx - \frac{R}{8\eta}  \frac{\partial \gamma}{\partial \bar{\mu}_R} \partial_\alpha \bar{\mu}_R \; ,
    \label{eq:dropletVelocity}
\end{equation}
where all quantities are evaluated at the droplet position $x^\mathrm{drop}_\alpha$.
In three dimensions, the prefactor $\frac18$ becomes $\frac{2}{15}$.
\Eqref{eq:dropletVelocity} clearly shows that larger droplets move faster, whereas higher viscosities~$\eta$ slow them down.
Additionally, the velocity is determined by the variation of surface tension and the chemical potential gradient. %

To test whether the prediction given by \Eqref{eq:dropletVelocity} is reasonable, we compare it to our numerical simulations.
We use the same setup as before (\secref{sec:numerical_simulations}) and run simulations of droplets with various radii~$R$ for different externally imposed gradients.
We determine droplet positions~$\vect{x}^\mathrm{drop}_i$ and radii~$R_i$ by fitting a spherical shape and determine the instantaneous  velocity $\vect{v}^\mathrm{drop}_i = (\vect{x}^\mathrm{drop}_{i+1} - \vect{x}^\mathrm{drop}_{i})/(t_{i+1} - t_i)$ for various time points $t_i$.
We then average the normalized velocity $\vect{v}^\mathrm{drop}_i / R_i$ for a short time period (after an initial transient) to estimate the normalized droplet velocity $\vect{v}^\mathrm{drop}/R$ (see Appendix \Figref{fig:Droplet_velocity_calc}).
To obtain a non-dimensional quantity, we report $\vect{v}^\mathrm{drop}/R$ in units of the time $\tau$ it takes for a particle to diffuse across the interface (\figref{fig:surfacetension}E).
Consequently, the observed low magnitudes are consistent with our assumption of quasi-stationarity.
The measured speed increases with the chemical potential difference $\Delta\bar\mu_R$ of the regulator between the two reservoirs, consistent with the last factor in \Eqref{eq:dropletVelocity}.
Similarly, the fact that data for different radii~$R$ collapses confirms the expected linear scaling with~$R$.
These observations confirm that \Eqref{eq:dropletVelocity} captures important qualitative aspects of droplet motion.

To quantitatively compare our numerical data to the prediction by \Eqref{eq:dropletVelocity}, we estimate $\partial\gamma/\partial\bar\mu_R$ and $\partial_\alpha\bar\mu_R$.
The first term follows directly by differentiating $\gamma(\bar\mu_R)$, which we obtained from the equilibrium considerations above (\figref{fig:surfacetension}B,C).
In contrast, the potential gradient $\partial_\alpha\bar\mu_R$ depends on the concrete compositions in the system.
\figref{fig:profiles}D suggests that $\bar\mu_R$ is a linear function of the position $x$ along the gradient both inside and outside the droplet, which is consistent with a stationary solution of the diffusion equation with weakly varying mobilities, $0 = \Lambda_\mathrm{eff}^\mathrm{in/out}\partial_\alpha^2 \bar\mu_R$.
To estimate the required gradient $\partial_\alpha\bar\mu_R$ inside the droplet, we consider a one-dimensional approximation, where $\bar\mu_R(x)$ is a piecewise linear function.
This function is uniquely defined given the overall chemical potential difference $\Delta\bar\mu_R$, system size $L$, droplet radius~$R$, and the respective mobilities $\Lambda_\mathrm{eff}^\mathrm{in/out}$, which we determine from $\Lambda_{ij}$ (see Appendix \secref{sec:estimation}).
Since $\Lambda_{ij}$ depends on densities, this procedure captures the physical effect that the potential gradient, and thus the droplet velocity, depend on the enrichment of the regulator in the droplet, which is controlled by the interaction matrix~$\chi_{ij}$.
Taken together, these ideas allow us to  use \Eqref{eq:dropletVelocity} to predict $\vect{v}^\mathrm{drop}$ without running a detailed simulation of \Eqsref{eqn:pde}--\eqref{eq:stokes}.
This prediction slightly overestimates the numerically measured values (\figref{fig:surfacetension}E), although it captures all trends in the numerical data, suggesting that the approximations that led to \Eqref{eq:dropletVelocity} capture the essence of the physical processes.
Interestingly, the analytical theory identifies a clear maximum in the velocity at intermediate chemical potentials~$\bar\mu_R$.

The maximal velocity at intermediate values of $\bar\mu_R$ originates from a non-monotonic dependence of $\partial\gamma/\partial\bar\mu_R$ on $\bar\mu_R$ (\figref{fig:surfacetension}C).
Such non-monotonic behavior is universal since $\partial\gamma/\partial\bar\mu_R$ must vanish for very small and very large values of $\bar\mu_R$:
On the one hand, the influence of the regulator must vanish in the binary limit ($\phi_R \to 0$, $\bar\mu_R/ \kT \to -\infty$).
On the other hand, the system is close to the critical point for large $\bar\mu_R$, where the surface tension~$\gamma$ exhibits the mean-field scaling $\gamma \sim |\bar{\mu}_R - \bar{\mu}_R^\mathrm{crit}|^{3/2}$ \cite{liuConcentrationDependenceInterfacial2012}.
Note that the scaling exponent is the same as for temperature $|T - T^\mathrm{crit}|$ \cite{cahnFreeEnergyNonuniform1958,pyoProximityCriticalityPredicts2023}, since the chemical potential difference scales proportional to the temperature difference to leading order as a consequence of field mixing \cite{sengersThermodynamicBehaviorFluids1986,fisherYangYangAnomalyFluid2000a}.
This scaling implies that $\partial\gamma/\partial\bar\mu_R$ vanishes for $\bar\mu_R\to\bar\mu_R^\mathrm{crit}$.
Since $\partial\gamma/\partial\bar\mu_R$ exhibits finite values while vanishing at extreme values of $\bar\mu_R$, its absolute value must assume a maximum, which implies a maximal speed~$v^\mathrm{drop}_\alpha$ given by \Eqref{eq:dropletVelocity}.
These properties do not depend on particular parameters, suggesting that knowledge of critical points is crucial to predict droplet propulsion.

\subsection{Morphology of phase diagrams determines the motion of droplets}
\label{sec:phase_diagram}

\begin{figure*}
    \centering
    \includegraphics[width=\textwidth]{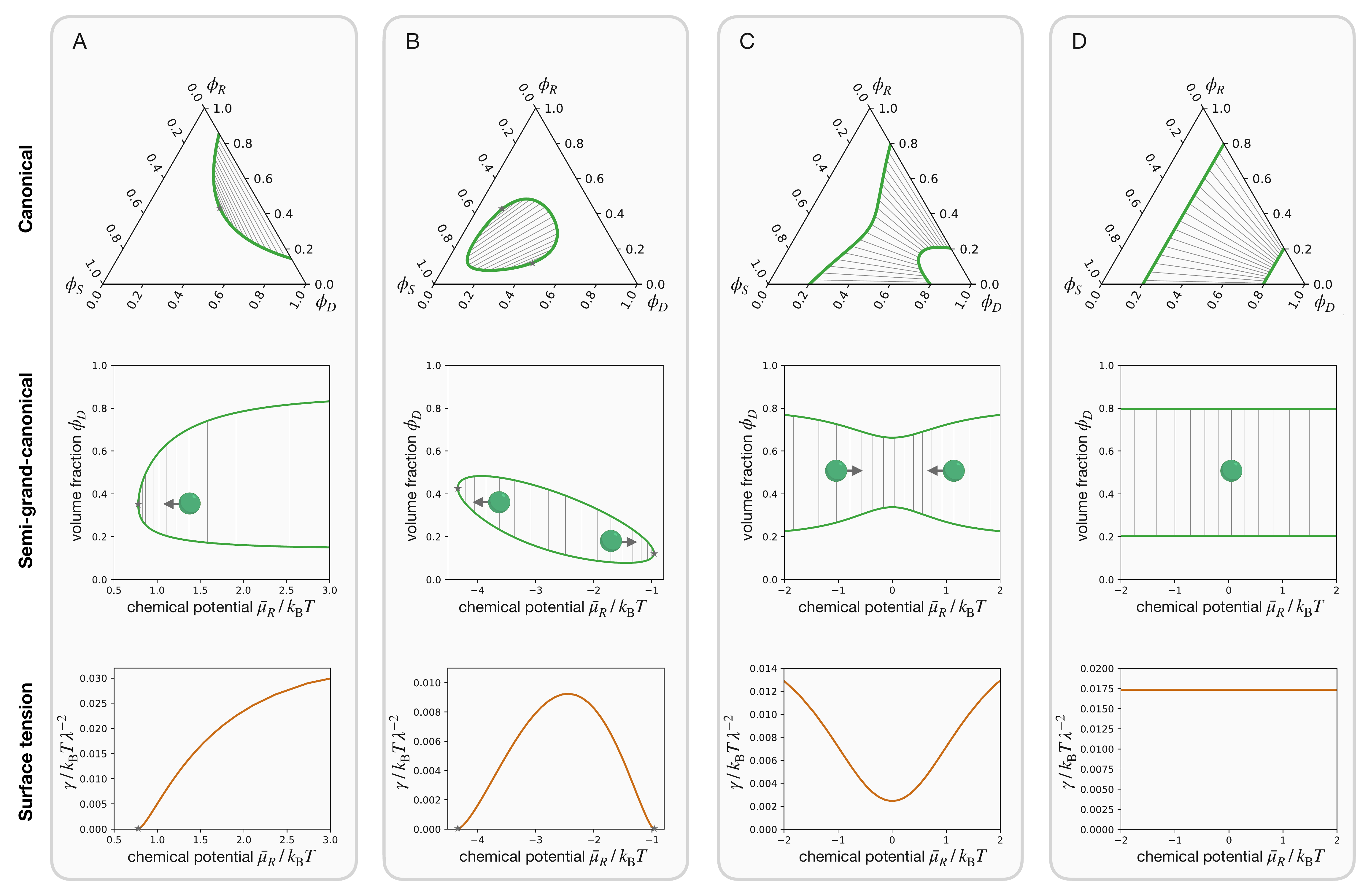}
    \caption{
        \textbf{Phase diagrams predict droplet motion.}
        Canonical phase diagram (upper row), grand-canonical phase diagram (middle row), and surface tension~$\gamma$ as a function of chemical potential $\bar\mu_R$ (lower row) of four qualitatively different cases (columns).
        Tie lines (gray) connect coexisting fractions of droplet material (green); Critical points are marked as stars.
        (A)~One critical point for $\chi_{DS} = 1.3$, $\chi_{DR} = 2.5$, and $\chi_{RS} = 1.7$. %
        (B)~Two critical points for $\chi_{DS} = 0.5$, $\chi_{DR} = -7$, and $\chi_{RS} = 0.5$. %
        (C)~No critical point for $\chi_{DS} = 2.3$, $\chi_{DR} = 2.3$, and $\chi_{RS} = 0.9$. %
        (D)~Marginal case without critical points for $\chi_{DS} = 2.3$, $\chi_{DR} = 2.3$, and $\chi_{RS} = 0$. %
        (A--D) $\kappa_{ij} = -\lambda^2(\chi_{ij} - \chi_{iS} - \chi_{Sj})$ for $i,j = D,R$, except in panel B, where $\kappa_{ij} = \bigl( \begin{smallmatrix} 0.75 & 0.375 \\ 0.375 & 0.75 \end{smallmatrix} \bigr)$.
    }
    \label{fig:phasediagrams}
\end{figure*}

To see whether the location of critical points is a robust predictor of droplet motion, we consider various phase diagrams with qualitatively different morphologies.
So far, we have considered a phase diagram in which increasing the regulator potential $\bar\mu_R$ destabilizes droplet formation (\figref{fig:surfacetension}A).
Since the regulator was dilute, this implied that the droplet moved toward regions of denser regulator (higher $\phi_R$).
In contrast, we could also consider systems in which the regulator stabilizes droplets, which implies a qualitatively different phase diagram (\figref{fig:phasediagrams}A).
The associated surface tension~$\gamma$ now increases with $\bar\mu_R$ (lower panel), such that \Eqref{eq:dropletVelocity} predicts that the droplet moves toward lower $\bar\mu_R$.
In both cases, the droplet thus moves toward the critical point, where surface tension vanishes.

Next, we consider more complex phase diagrams.
\Figref{fig:phasediagrams}B shows a case of segregative phase separation between droplet material and regulator, which exhibits two critical points.
Since this case involves strong attraction ($\chi_{DR}=-7$), the corresponding $\kappa_{ij} = -\lambda^2(\chi_{ij} - \chi_{iS} - \chi_{Sj})$ would not be positive definite, and we thus choose $\kappa_{ij}$ independent of $\chi_{ij}$ in this case.
In any case, the surface tension vanishes at the two critical points, implying the associated~$\gamma$ depends non-monotonically on $\bar\mu_R$ (lower panel).
The direction of droplet propulsion depends on $\bar\mu_R$: For low $\bar\mu_R$, the droplet moves toward the left critical point, whereas for large $\bar\mu_R$, it moves toward the right one.
Consequently, in the same gradient, droplets can move in opposite directions depending on the location along the gradient.

We also note the existence of phase diagrams without critical points:
\Figref{fig:phasediagrams}C shows a generic example where the droplet material segregates from solvent and regulator.
Here, $\gamma$ also exhibits a non-monotonic dependence on $\bar\mu_R$, although with the opposite curvature compared to panel~B.
Consequently, droplets move toward the center of the phase diagram, where $\gamma$ is minimal.
Between these two cases is a marginal case (\figref{fig:phasediagrams}D), where solvent and regulator have identical physical properties ($\chi_{DS}=\chi_{DR}$ and $\chi_{RS}=0$).
This case is essentially binary phase separation between droplet material and the other species, implying that the phase diagram and $\gamma$ do not depend on the relative fractions of solvent to regulator, and externally imposed regulator gradients do not induce propulsion. 
Note that this result relies on the connection between $\kappa_{ij}$ and $\chi_{ij}$ described in \secref{sec:numerical_simulations}.
This relation is not necessarily accurate for complex molecules, which could lead to surface tension gradients even when the phase diagram predicts none (see Appendix \secref{sec:SIkappa}).
However, since surface tensions must vanish at critical points, these details have little influence on the qualitative conclusion that the morphology of the phase diagram shapes the propulsion of the droplet. 

\begin{figure}
    \centering
    \includegraphics[width=0.5\textwidth]{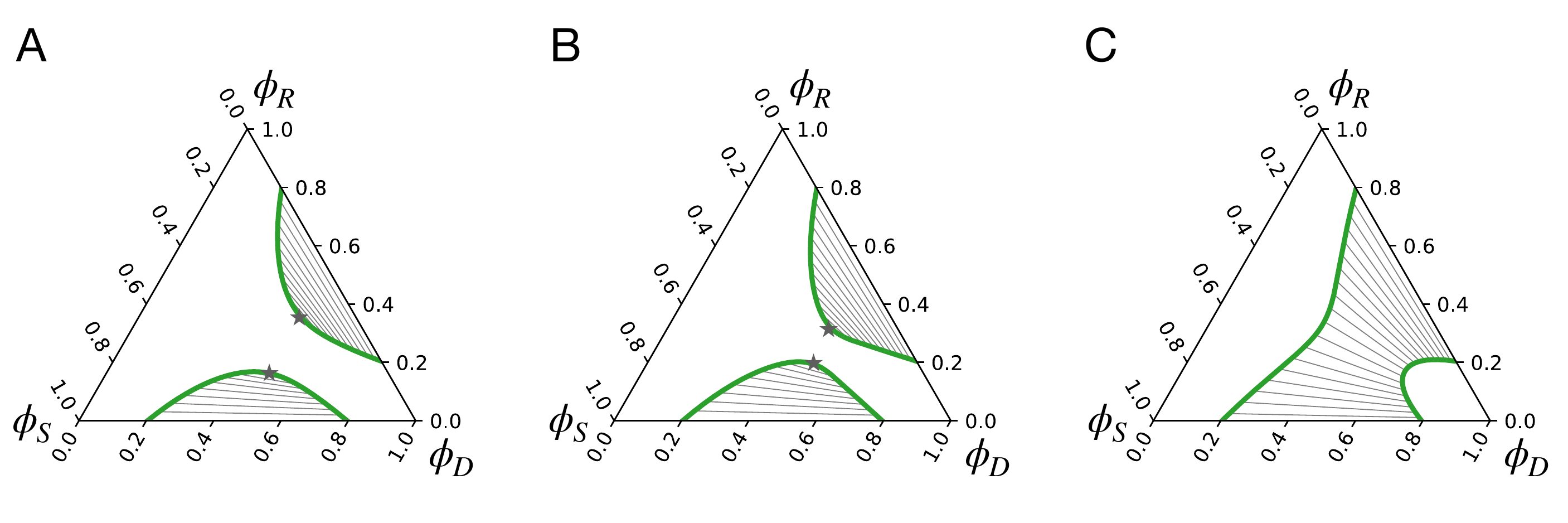}
    \caption{
        \textbf{Similarity of phase diagrams informs droplet motion in absence of critical points.}
        Canonical phase diagrams at $\chi_{DS} = \chi_{DR} = 2.3$ and $\chi_{RS} = \{1.5, 1.3, 0.9\}$ (left-to-right).
        Tie lines (gray) connect coexisting fractions of droplet material (green); Critical points are marked by stars.
    }
    \label{fig:phasediagram_morph}
\end{figure}

In the simplest case, the morphology of phase diagrams is controlled by critical points (\figref{fig:phasediagrams}A,B), and the droplet generally moves toward those.
However, even if the phase diagram does not possess critical points (\figref{fig:phasediagrams}C), we can use qualitative information on how phase diagrams depend on interaction parameters to deduce the direction of droplet propulsion:
The particular phase diagram shown in \figref{fig:phasediagrams}C is closely related to a case with two 2-phase regions, where droplet material segregates either from solvent or regulator, while these two species repel each other (large $\chi_{RS}$, \figref{fig:phasediagram_morph}A).
The corresponding critical points are close to the center of the phase diagram, implying that droplets would move toward the center in either 2-phase region.
Decreasing $\chi_{RS}$ morphs the phase diagram toward the one shown \figref{fig:phasediagrams}C, demonstrating that they are closely related.
In particular, the phase diagram inherits the direction of droplet propulsion, even though the critical points disappeared.
This behavior persists until $\chi_{RS}$ vanishes at the marginal case (\figref{fig:phasediagrams}D).
Taken together, the presence of critical points at nearby parameter values informs the phase diagram morphology and thus droplet propulsion in chemical gradients.
As a rule-of-thumb, droplets perform \emph{dialytaxis}~\cite{jambon-puilletPhaseseparatedDropletsSwim2024}, i.e., they move toward their dissolution.

\section{Discussion}\label{sec:discussion}

In this work, we have described surface-tension-driven propulsion of a droplet in a regulator gradient.
In contrast to previous literature~\cite{youngMotionBubblesVertical1959,ratkeInfluenceParticleMotion1985,tegzePhaseFieldSimulation2005,schmittMarangoniFlowDroplet2016,yabunakaSelfpropelledMotionFluid2012,yoshinagaSpontaneousMotionDeformation2014}, surface tension is not prescribed, but emerges from the interactions of droplet components, solvent, and regulator, which are all treated as continuous fields. 
We found that this situation is best described in a semi-grand-canonical ensemble, where the surface tension is evaluated as a function of the chemical potential of the regulator.
A regulator gradient then implies a surface tension gradient, which propels the droplet via the Marangoni effect.
We derived a compact expression for the droplet velocity $v^\mathrm{drop}_\alpha$ in \Eqref{eq:dropletVelocity}, showing that,
qualitatively, $v^\mathrm{drop}_\alpha$ is determined by the equilibrium phase diagram via $\partial\gamma/\partial\bar\mu_R$ (\figref{fig:phasediagrams}).
Quantitatively, $v^\mathrm{drop}_\alpha$ can be affected by multiple parameters:
First, by the viscosity~$\eta$ of the mixture.
Second, by droplet size~$R$, which is determined by the total amount of droplet material.
Third, by $\partial\gamma/\partial\bar\mu_R$, whose magnitude is governed by the average surface tension and whose form is determined by the interaction matrix~$\chi_{ij}$, and thus tightly linked with the phase diagram.
Fourth, by the non-equilibrium drive in the form of $\partial_\alpha\bar\mu_R$, which is controlled by the externally applied gradient and modified by the partition coefficient of the regulator.

To estimate the propulsion speed $v^\mathrm{drop}$ quantitatively, we consider two example experimental systems.
The first system comprises P granules that segregate in \mbox{MEX-5} gradients in \textit{C.~elegans} embryos~\cite{brangwynneGermlineGranulesAre2009}.
These granules have typical radii of $\SI{0.3}{\micro\meter}$ and exhibit relatively large ratios of viscosity to surface tension, $\eta/\gamma \sim \SI{2}{\second\per\micro\meter}$, possibly because the complex interface implies small~$\gamma$~\cite{Folkmann2021}.
To estimate the maximal velocity, we assume that surface tension vanishes on one side of the cell of length $L\sim\SI{50}{\micro\meter}$, implying $v^\mathrm{drop}\sim\SI{0.4}{\nano\meter\per\second}$.
This velocity is small (e.g., compared to flows in the embryo on the order of $v \sim \SI{50}{\nano\meter\per\second}$~\cite{brangwynneGermlineGranulesAre2009}), so surface-tension-driven propulsion is negligible in P granule segregation.
The second system concerns the strong self-propulsion of BSA droplets in pH gradients~\cite{jambon-puilletPhaseseparatedDropletsSwim2024}.
Here, typical parameters are
$R \sim \SI{50}{\micro\meter}$, 
$\eta \sim \SI{0.5}{\pascal\second}$, 
$\eta/\gamma\sim \SI{5e-3}{\second\per\micro\meter}$,
$\partial \gamma/\partial \mathrm{pH} \sim -40/\mathrm{pH} \ \SI{}{\micro\newton\per\meter}$, and
$\partial \mathrm{pH}/\partial x\sim\SI{7e-4}{pH \per\micro\meter}$,
leading to
$v^\mathrm{drop} \sim \SI{0.4}{\micro\meter\per\second}$, consistent with measurements~\cite{jambon-puilletPhaseseparatedDropletsSwim2024}.
These examples show that the \textit{in vitro} BSA system exhibits velocities that are roughly three orders of magnitude larger than those in the \textit{in vivo} P granule system.
While the \textit{in vitro} system exhibits much smaller ratios $\eta/\gamma$, gradients are also applied over larger distances, so that these two effects roughly compensate.
The velocity difference is thus mainly due to the size disparity of droplets.

So far, we have only considered the simplest case of an isolated droplet in a single regulator gradient in the limit where diffusion-based drift is negligible~\cite{weberDropletRipeningConcentration2017,sahaPolarPositioningPhaseSeparated2016,hafnerReactionDrivenDiffusiophoresisLiquid2024}.
Since drift can lead to self-propulsion~\cite{demarchiEnzymeEnrichedCondensatesShow2023} and interactions between droplets~\cite{ goychukAnomalousDiffusionDirected2025}, we expect similar effects from hydrodynamic propulsion, potentially including nonreciprocal interactions~\cite{meredithPredatorPreyInteractions2020}.
Furthermore, our theory can be readily generalized to multiple chemical potential gradients in $N$-component systems with a single droplet forming component.
Moreover, propulsion in realistic droplets might be affected by viscoelasticity~\cite{jawerthSaltDependentRheologySurface2018,styleLiquidLiquidPhaseSeparation2018,jawerthProteinCondensatesAging2020,rosowskiElasticRipeningInhibition2020, paulinActiveViscoelasticCondensates2025} and surfactants~\cite{maassSwimmingDroplets2016,hardyKineticTheoryCoupled2025,thutupalliSwarmingBehaviorSimple2011,michelinSpontaneousAutophoreticMotion2013,herminghausInterfacialMechanismsActive2014,jinChemotaxisAutochemotaxisSelfpropelling2017,vossChemomechanicalMotilityModes2025}.
Finally, self-generated gradients, e.g., from evaporation~\cite{lohsePhysicochemicalHydrodynamicsDroplets2020,diddensCompetingMarangoniRayleigh2021} or chemical reactions~\cite{jambon-puilletPhaseseparatedDropletsSwim2024,dindoChemotacticInteractionsDrive2024,kostlerAdvectionSelectsPattern2026}, could lead to additional dynamical phenomena.

\begin{acknowledgments}
We thank Gregor Häfner, Marcus Müller, Oliver Paulin, Kaarthik Varma, and Eric R. Dufresne for fruitful discussions.
We gratefully acknowledge funding from the Max Planck Society. S. K. acknowledges funding from a fellowship of the IMPRS-PBCS. S. K. acknowledges support by the study program “Biological Physics” of the Elitenetzwerk Bayern.
\end{acknowledgments}

\bibliography{Bibliography.bib, ManualBib.bib}

\clearpage %
\onecolumngrid  %

\renewcommand\thefigure{S\arabic{figure}}
\setcounter{figure}{0}

\renewcommand{\theequation}{\thesection.\arabic{equation}}
\counterwithin*{equation}{section}
\setcounter{equation}{0}

\appendix

\section{Numerical solution of continuous equations}
\label{sec:SInumerics}

To solve the dynamics of the volume fractions $\phi_i$ (\Eqref{eqn:pde} in the main text) numerically, we use a finite difference method with Euler time-stepping implemented in the \href{https://github.com/zwicker-group/py-pde}{\textit{py-pde}} Python package~\cite{Zwicker2020}.
In particular, we solve the Stokes equation (\Eqref{eq:stokes} in the main text), which governs the center-of-mass velocity field $v_\alpha$ for each density field~$\phi_i$,
\begin{align}
    \eta \partial_\alpha^2 v_\beta - \partial_\beta p &= - f_\beta
    &
    \partial_\alpha v_\alpha &= 0 \; .
\end{align}
We solve these equations by applying the Oseen propagator to the forcing $f_\beta = \partial_\alpha \sigma_{\alpha\beta} = - \sum_i \nu_i^{-1} \phi_i \partial_\beta \bar\mu_i$ in Fourier space.
Taking a divergence and using $\partial_\beta v_\beta = 0$, we get
\begin{equation}
    \partial_\beta^2 p = \partial_\beta f_\beta
    \;.
\end{equation}
Fourier transforming both equations leads to
\begin{align}
    \hat{v}_\alpha &= - \frac{ik_\alpha \hat{p} - \hat{f}_\alpha}{\eta \, k^2}
& \text{and} &&
    \hat{p} &= - \frac{ik_\alpha \hat{f}_\alpha}{k^2}
    \;.
\end{align}
Plugging the second expression into the first, we end up with
\begin{equation}
    \hat{v}_\alpha = \frac{\hat{f}_\alpha}{\eta \, k^2} - \frac{k_\alpha (k_\beta \hat{f}_\beta)}{\eta \, k^4} = P_{\alpha\beta} \hat{f}_\beta
    \;,
\end{equation}
where 
\begin{equation}
    P_{\alpha\beta} = \frac{1}{\eta} \left( \frac{\delta_{\alpha\beta}}{k^2} - \frac{k_\alpha k_\beta}{k^4}\right)
\end{equation}
is the Oseen propagator, which essentially projects the force to a divergence free (solenoidal) subspace.

In our concrete case, we solve the problem on a periodic domain, which allows using fast Fourier transforms to evaluate the Oseen propagator efficiently.
The reservoir boundary conditions $\bar\mu_R(x_+) = \bar\mu_R^+$ and $\bar\mu_R(x_-) = \bar\mu_R^-$, the no-flux boundary conditions for the droplet material, and the equal-interaction conditions $\partial_n \phi_i = 0$ are imposed at virtual boundaries inside the periodic domain at positions $x_\pm$ inside the periodic domain.
Note that we choose the system large enough, so that the velocity field $v_\alpha$ has sufficiently decayed at the virtual boundaries and the periodic boundary conditions do not affect the droplet dynamics.
Throughout the paper we set the interfacial width $\lambda = 0.5$, the mobility $\Lambda_0 = 1$, the molecular volumes $\nu_i = \nu_0 = 1$ for all $i$, and the energy scale $k_\mathrm{B} T = 1$.

\section{Derivation of local stress interface conditions}
\label{sec:SIstressinterface}

\begin{figure}
    \centering
    \includegraphics[width=\linewidth]{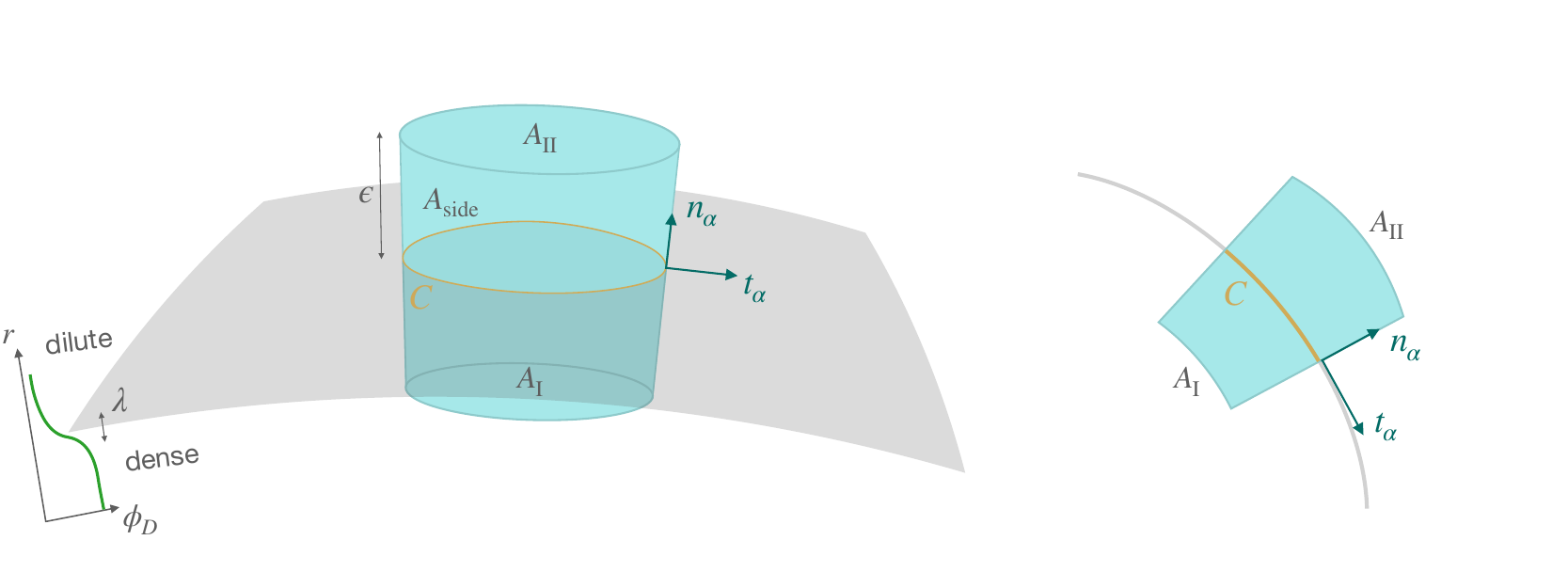}
    \caption{
        \textbf{Schematic of local volume at interface} 
        showing the interface (gray) defined as an isocontour of the droplet component $\phi_D$ (green schematic plot on the left; $\lambda$ denotes the interfacial width). 
        The cylinder volume (teal) to derive the local stress balance at the interface extends a distance $\epsilon$ in the normal direction $n_\alpha$ into both phases and intersects the interface at a curve $C$ (yellow). 
        The outward normal of the cylinder side surface $A_\mathrm{side}$ lies in the local tangent plane of the interface.
        The schematic on the right shows the analogous situation for two dimensions.
    }
    \label{fig:schematic}
\end{figure}

\begin{figure}
    \centering
    \includegraphics[width=\linewidth]{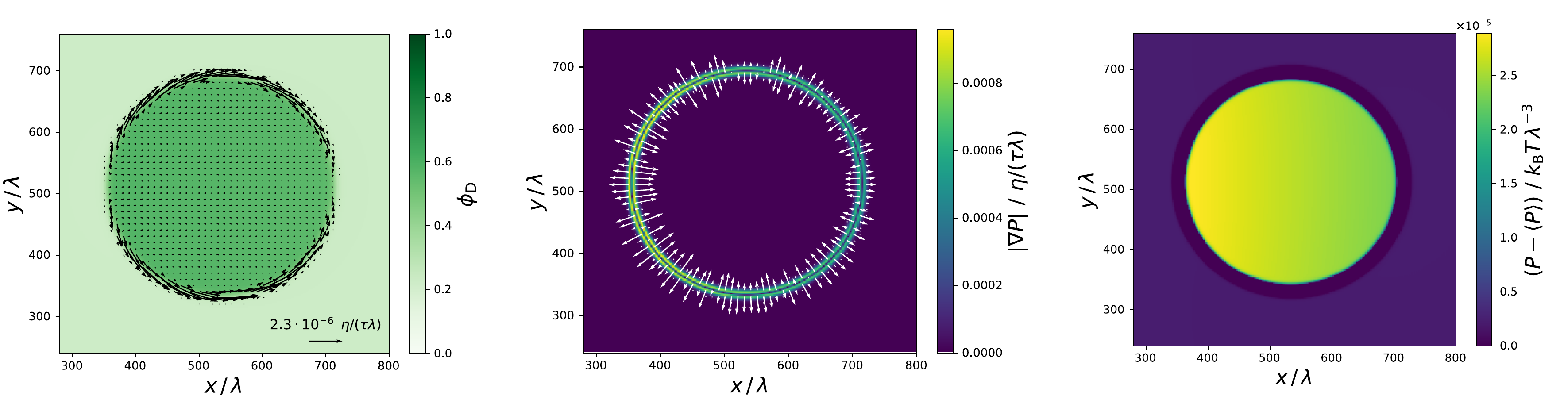}
    \caption{
        \textbf{Force densities resulting from viscous stress and total pressure.}
        (left) Divergence of viscous stress $\partial_\alpha (\eta e_{\alpha\beta})$ (black arrows) together with volume fraction~$\phi_D$ (color scale).
        (middle) Gradient of total pressure $P$ (arrows) and its magnitude (color) in units of $\eta/(\tau\lambda)$ for comparison with $\partial_\alpha (\eta e_{\alpha\beta})$.
        (right) Total pressure deviation from average pressure $\langle P \rangle$.
        In all panels, model parameters are $R_0 = 180 \lambda$, $\Delta \bar{\mu}_R = 0.15 \, \kT$, $\bar\mu_R^+ = - 0.9 \, \kT$, $\chi_{DS} = 2.5$, $\chi_{DR} = 1.3$, $\chi_{RS} = 1.7$,
        $\tau = \lambda^2/(\Lambda_0 \ \kT)$, $\Lambda_{ij} = \Lambda_0 (\phi_i \delta_{ij} - \phi_i\phi_j)$,
        $L_x = 984 \, \lambda$, and $L_y = 1024 \, \lambda$.
    }
    \label{fig:force_densities}
\end{figure}

We derive the thin interface condition for the stress across the interface (\Eqref{eqn:stokes_bcs} in the main text) by following the approach of \refcite{andersonDIFFUSEINTERFACEMETHODSFLUID1998}.
Focusing on a small cylinder across the interface (teal structure on the left in \figref{fig:schematic}), we can integrate the local stress balance (\Eqref{eq:stokes} in the main text) and use the divergence theorem to find the jump conditions for the stress across the interface.
The upper and lower surfaces of the cylinder at $-\epsilon$ and $\epsilon$ are located in the bulk phases I and II,  while the side surface covers the interfacial region and its normal $t_\alpha$ is tangential to the interface (\figref{fig:schematic} left).
We assume that the cylinder is tall compared to the interfacial width $\lambda$ ($\epsilon \gg \lambda$), so that the interface is fully contained.
At the same time, we assume that $\epsilon$ is much smaller than the length scale of compositional gradients inside the phases, such that $\tau_{\alpha\beta}$ vanishes at the upper and lower surfaces, which we denote by  $A_\mathrm{II}$  and  $A_\mathrm{I}$, respectively.
Furthermore, we assume that the curvature of the interface is much smaller than $1/\epsilon$ ($\epsilon H \ll 1$, where $H$ is the sum of principal curvatures), and that effects of Gaussian curvature are negligible.
Integrating the stress balance over the cylinder volume and applying the divergence theorem, we get
\begin{align}
    0 =& \ \int_{A_\mathrm{II}} \diff A \ n_\alpha \left[ \eta\left( \partial_\alpha v_\beta + \partial_\beta v_\alpha \right) - \delta_{\alpha\beta} \, p - \delta_{\alpha\beta} \, \Pi\right] \notag \\
    &- \int_{A_\mathrm{I}} \diff A \ n_\alpha \left[ \eta\left( \partial_\alpha v_\beta + \partial_\beta v_\alpha \right) - \delta_{\alpha\beta} \, p - \delta_{\alpha\beta} \, \Pi\right] \notag \\
    &+ \int_{A_\mathrm{side}} \diff A \ t_\alpha \left[ \eta\left( \partial_\alpha v_\beta + \partial_\beta v_\alpha \right) - \delta_{\alpha\beta} \, p + \sigma_{\alpha\beta}\right]
\end{align}
Here, $n_\alpha$ the outward normal to the interface, defined as the normal to an isocontour of a singled out droplet component $\phi_D$.
The cylinder is oriented such that $n_\alpha$ is also parallel to the normal of the upper and lower surfaces. 
In contrast, the normal to the side surface coincides with the tangent to the interface $t_\alpha$, with $n_\alpha t_\alpha = 0$.
For a sufficiently small cylinder volume, the rate-of-strain terms in the last integral over the side surface cancel each other (assuming a sufficiently smooth velocity field).
Combining the hydrostatic pressure~$p$ and osmotic pressure~$\Pi$ into the total pressure $P = p + \Pi$, we thus get
\begin{align}
    0 =& \ \int_{A_\mathrm{II}} \diff A \ n_\alpha \left[ \eta\left( \partial_\alpha v_\beta + \partial_\beta v_\alpha \right) - \delta_{\alpha\beta} \, P\right] \notag \\
    &- \int_{A_\mathrm{I}} \diff A \ n_\alpha \left[ \eta\left( \partial_\alpha v_\beta + \partial_\beta v_\alpha \right) - \delta_{\alpha\beta} \,P\right] \notag \\
    &+ \int_{A_\mathrm{side}} \diff A \ t_\alpha \left[ - \delta_{\alpha\beta} \, P + \tau_{\alpha\beta}\right]
\end{align}
Since $\partial_\alpha \phi_i \approx n_\alpha \partial_n \phi_i$, the normal components dominate the capillary stress ($\tau_{nn}, \, \Pi \gg \tau_{nt}, \, \tau_{tt}$), implying that the integral over the side becomes
\begin{align}
    \int_{A_\mathrm{side}} \diff A \ t_\alpha \left[ - \delta_{\alpha\beta} \, P + \tau_{\alpha\beta}\right] \approx \oint_{C} \diff s \int_{-\epsilon}^\epsilon \diff r \ t_\alpha \left[- P + \tau_{tt}\right] \approx -\oint_{C} \diff s \int_{-\epsilon}^\epsilon \diff r \ t_\alpha P
    \;,
    \label{eq:pressure_integral}
\end{align}
where we decomposed the integration of the side surface into an integration in the normal direction $n_\alpha$ and an integration along the contour $C$ on the surface. 
To evaluate the integral $-\int \diff r P$, we use the definition of the chemical potential,
\begin{equation}
    \frac{\bar\mu_i}{\nu_i} = \frac{\kT}{\nu_0} \left( \frac{\partial f_0}{\partial \phi_i} - \sum_j \kappa_{ij} \partial_\alpha^2 \phi_j \right) \; .
\end{equation}
Multiplying both sides with $\partial_n \phi_i$, summing over all components and integrating along the direction normal to the droplet interface, we get
\begin{equation}
    \int_{r_0}^{r'} \diff r \ \sum_i \frac{\bar\mu_i}{\nu_i} \partial_n \phi_i = \frac{\kT}{\nu_0} \int_{r_0}^{r'} \diff r \  \sum_i \left( \frac{\partial f_0}{\partial \phi_i} - \sum_j \kappa_{ij} \partial_\alpha^2 \phi_j \right) \partial_n \phi_i \; ,
\end{equation}
where $r_0$ lies in the bulk phase.
Using partial integration, we find
\begin{equation}
    \int_{r_0}^{r'} \diff r \ \partial_n \sum_i \frac{\bar\mu_i}{\nu_i} \phi_i - \int_{r_0}^{r'} \diff r \  \sum_i \frac{\phi_i}{\nu_i} \partial_n \bar\mu_i = \frac{\kT}{\nu_0} \int_{r_0}^{r'} \diff r \ \left( \partial_n f_0 - \sum_{ij} \kappa_{ij} (\partial_\alpha^2 \phi_j) (\partial_n \phi_i) \right) \; ,
\end{equation}
where we also used the chain rule $\partial_n f_0 = \sum_i \frac{\partial f_0}{\partial \phi_i} \partial_n \phi_i$. Approximating $\partial_\alpha = n_\alpha \partial_n$ in the interfacial region, we obtain $\partial_\alpha^2 = \partial_n^2 + H \partial_n$, with $H = \partial_\alpha n_\alpha$ the sum of principal curvatures.
Therefore,
\begin{align}
    \int_{r_0}^{r'} \diff r \ \partial_n \sum_i \frac{\bar\mu_i}{\nu_i} \phi_i - \int_{r_0}^{r'} \diff r \  \sum_i \frac{\phi_i}{\nu_i} \partial_n \bar\mu_i &= \frac{\kT}{\nu_0} \int_{r_0}^{r'} \diff r \ \left( \partial_n f_0 - \sum_{ij} \kappa_{ij} (\partial_n^2 \phi_j) (\partial_n \phi_i) - \sum_{ij} \kappa_{ij} H (\partial_n \phi_j) (\partial_n \phi_i) \right) 
\notag\\
    &= \frac{\kT}{\nu_0} \int_{r_0}^{r'} \diff r \ \left( \partial_n f_0 - \frac{1}{2} \partial_n \sum_{ij} \kappa_{ij} (\partial_n \phi_i) (\partial_n \phi_j) - \sum_{ij} \kappa_{ij} H (\partial_n \phi_i) (\partial_n \phi_j) \right) 
\notag\\
    &= \int_{r_0}^{r'} \diff r \ \left( \partial_n f - \frac{\kT}{\nu_0} \partial_n \sum_{ij} \kappa_{ij} (\partial_n \phi_i) (\partial_n \phi_j) - \frac{\kT}{\nu_0} \sum_{ij} \kappa_{ij} H (\partial_n \phi_i) (\partial_n \phi_j) \right) \; .
\end{align}
Using the definition of the osmotic pressure, $\Pi = \sum_i \nu_i^{-1} \phi_i \bar\mu_i - f$, we find
\begin{equation}
    \Pi(r') - \Pi(r_0) + \int_{r_0}^{r'} \diff r \  \partial_n p =  -  \frac{\kT}{\nu_0} \sum_{ij} \kappa_{ij} (\partial_n \phi_i) (\partial_n \phi_j) - \frac{\kT}{\nu_0} \int_{r_0}^{r'} \diff r \ \sum_{ij} \kappa_{ij} H (\partial_n \phi_i) (\partial_n \phi_j) \; ,
\end{equation}
where, we use force balance $ - \sum_i \phi_i \nu_i^{-1} \partial_\beta \bar\mu_i = \partial_\alpha \sigma_{\alpha\beta} = \partial_\beta p$, to simplify the second term on the left hand side.
Here, we assumed negligible viscous stress, which is substantiated by numerical simulations (\Figref{fig:force_densities}), that show that $\partial_\alpha (\eta e_{\alpha\beta})$ is two to three orders of magnitude smaller than contributions from total pressure.
Furthermore, $\partial_\alpha (\eta e_{\alpha\beta})$ is oriented tangentially to the interface, so its normal projection will vanish.
Therefore, the left hand side can be expressed in terms of the total pressure $P = \Pi + p$,
\begin{equation}
    P(r') - P(r_0) =  -  \frac{\kT}{\nu_0} \sum_{ij} \kappa_{ij} (\partial_n \phi_i) (\partial_n \phi_j) - \frac{\kT}{\nu_0} \int_{r_0}^{r'} \diff r \ \sum_{ij} \kappa_{ij} H (\partial_n \phi_i) (\partial_n \phi_j) \; .
\end{equation}
Since pressures are only defined up to a constant, we set $P(r_0) = 0$ as a reference pressure.
Integration over the interface from $-\epsilon$ to $\epsilon$ gives
\begin{equation}
    \int_{-\epsilon}^{\epsilon} \diff r' \ P =  - \gamma - \frac{\kT}{\nu_0} \int_{-\epsilon}^{\epsilon} \diff r' \ \int_{r_0}^{r'} \diff r \ \sum_{ij} \kappa_{ij} H (\partial_n \phi_i) (\partial_n \phi_j) \; .
\end{equation}
In the last term, the integration can be broken up and the order of integration can be exchanged, resulting in 
\begin{equation}
     -\frac{\kT}{\nu_0} \int_{r_0}^{-\epsilon} \diff r \int_{-\epsilon}^{\epsilon} \diff r' \ H \sum_{ij} \kappa_{ij} (\partial_n \phi_i) (\partial_n \phi_j) -\frac{\kT}{\nu_0} \int_{-\epsilon}^{\epsilon} \diff r \int_{r}^{\epsilon} \diff r' \ H \sum_{ij} \kappa_{ij} (\partial_n \phi_i) (\partial_n \phi_j) \; .
\end{equation}
Since the expressions depend on $r$ but not $r'$ we can perform the $r'$ integration,
\begin{equation}
     -\frac{\kT}{\nu_0} \int_{r_0}^{-\epsilon} \diff r \ 2\epsilon H \sum_{ij} \kappa_{ij} (\partial_n \phi_i) (\partial_n \phi_j) -\frac{\kT}{\nu_0} \int_{-\epsilon}^{\epsilon} \diff r \ (\epsilon - r) H \sum_{ij} \kappa_{ij} (\partial_n \phi_i) (\partial_n \phi_j) \; .
\end{equation}
The first integral is much smaller than the second one, since $\partial_n \phi_i$ is negligible for $r$ outside $[-\epsilon, \epsilon]$.
Splitting the second integral into two parts, we get
\begin{equation}
    -\frac{\kT}{\nu_0} \int_{-\epsilon}^{\epsilon} \diff r \ \epsilon H \sum_{ij} \kappa_{ij} (\partial_n \phi_i) (\partial_n \phi_j) + \frac{\kT}{\nu_0} \int_{-\epsilon}^{\epsilon} \diff r \ r H \sum_{ij} \kappa_{ij} (\partial_n \phi_i) (\partial_n \phi_j) \; .
\end{equation}
Assuming that $H$ does not change significantly in the $\epsilon$ interval, and using that $\sum_{ij} \kappa_{ij} (\partial_n \phi_i) (\partial_n \phi_j)$ is a positive quantity for physically meaningful $\kappa_{ij}$ (positive definite), the magnitude of this total expression can be bounded by $2 \epsilon H \gamma$.
In the thin-interface limit ($\epsilon H \ll 1$) we neglect this contribution to arrive at the final expression
\begin{equation}
    \gamma \approx - \int_{-\epsilon}^{\epsilon} \diff r' \ P \; .
\end{equation} 
Note that although the line integral of $P$ depends on the arbitrary pressure offset, only differences along the interface enter the side-force balance.
A constant shift of $P$ contributes equally along the cylinder contour and therefore does not affect the resulting surface-gradient force.
For a sufficiently small cylinder, the two surfaces have similar areas, $A_\mathrm{I} \approx A_\mathrm{II} = A$, implying
\begin{align}
    0 = \int_{A} \diff A \ n_\alpha \left[ \eta\left( \partial_\alpha v_\beta + \partial_\beta v_\alpha \right) - \delta_{\alpha\beta} \, P\right]^\mathrm{II}_\mathrm{I} + \oint_{C} \diff s \ t_\alpha \gamma
    \;.
\end{align}
Even though the osmotic pressure can be absorbed into the hydrostatic pressure, and thus seems not to contribute to flows, it gives rise to the surface tension gradient $\nabla^S_\beta \gamma$.
Using the surface divergence theorem \cite{scrivenDynamicsFluidInterface1960},  we find
\begin{align}
    0 = \int_{A} \diff A \left\{ n_\alpha \Big[ \eta\left( \partial_\alpha v_\beta + \partial_\beta v_\alpha \right) - \delta_{\alpha\beta} \, P\Big]^\mathrm{II}_\mathrm{I} + \nabla^S_\beta \gamma - H \gamma n_\beta \right\}
    \;,
\end{align}
where we again used that the area enclosed by curve $C$ is approximately $A$ for a thin cylinder.
This condition provides boundary conditions given by \Eqsref{eqn:stokes_bcs} on the stress at the interface, while the stress balance in the bulk phases is given by
\begin{equation}
    \partial_\alpha \bigl[ \eta\left( \partial_\alpha v_\beta + \partial_\beta v_\alpha \right) - \delta_{\alpha\beta} \, P\bigr] = 0 \; .
\end{equation}
In two dimensions, the integral over the sides results in
\begin{equation}
    - \int_{-\epsilon}^{\epsilon} \diff r \ t_\beta P + \int_{-\epsilon}^\epsilon \diff r \ t'_\beta P = t_\beta \gamma - t'_\beta \gamma \; ,
\end{equation}
where $t_\alpha$ and $t'_\alpha$ are tangent vectors at the two sides of the 2D integration area (teal structure on the right in \Figref{fig:schematic}).
For a thin integration area, we find
\begin{equation}
    t_\beta \gamma - t'_\beta \gamma = \int_C \diff s \ \frac{\partial (t_\beta \gamma)}{\partial s} \; ,
\end{equation}
where $s$ is the arc-length along the interface.
We get
\begin{equation}
    \int_C \diff s \left(\frac{\partial t_\beta}{\partial s} \gamma + t_\beta \frac{\partial \gamma}{\partial s} \right) = \int_C \diff s \left(-H n_\beta \gamma + \nabla^S_\beta \gamma \right) \; ,
\end{equation}
which leads again to the total expression
\begin{equation}
    0 = \int_C \diff s \left(n_\alpha \Big[ \eta\left( \partial_\alpha v_\beta + \partial_\beta v_\alpha \right) - \delta_{\alpha\beta} \, P \Big]_\mathrm{I}^\mathrm{II}  + \nabla^S_\beta \gamma-H n_\beta \gamma\right) \; .
\end{equation}

\section{Derivation of droplet velocity}
We here briefly trace the derivation of the droplet velocity given by \Eqref{eq:dropletVelocity} from \Eqref{eqn:drop_speed}.
Starting from \Eqref{eqn:drop_speed}, we expand the surface tension to first order in the variation of $\bar\mu_R$ across the droplet,
\begin{equation}
    \gamma(\theta) \approx \gamma(\bar\mu_R) + \frac{\partial \gamma}{\partial \bar{\mu}_R} \, R  \, \hat{r}_\beta \ \partial_\beta \bar{\mu}_R \; ,
\end{equation}
where $\bar\mu_R$, $\frac{\partial \gamma}{\partial \bar{\mu}_R}$ and $\partial_\alpha \bar\mu_R$ are evaluated at the droplet center.
Using $\int d\Omega \ \hat{r}_\alpha \hat{r}_\beta = k_d \pi \delta_{\alpha\beta}$, with $k_2 = 1$ for 2D and $k_3 = \frac{4}{3}$ for 3D, we arrive at \Eqref{eq:dropletVelocity}.

\section{Numerical evaluation of normalized droplet velocity}

\begin{figure}
    \centering
    \includegraphics[width=\linewidth]{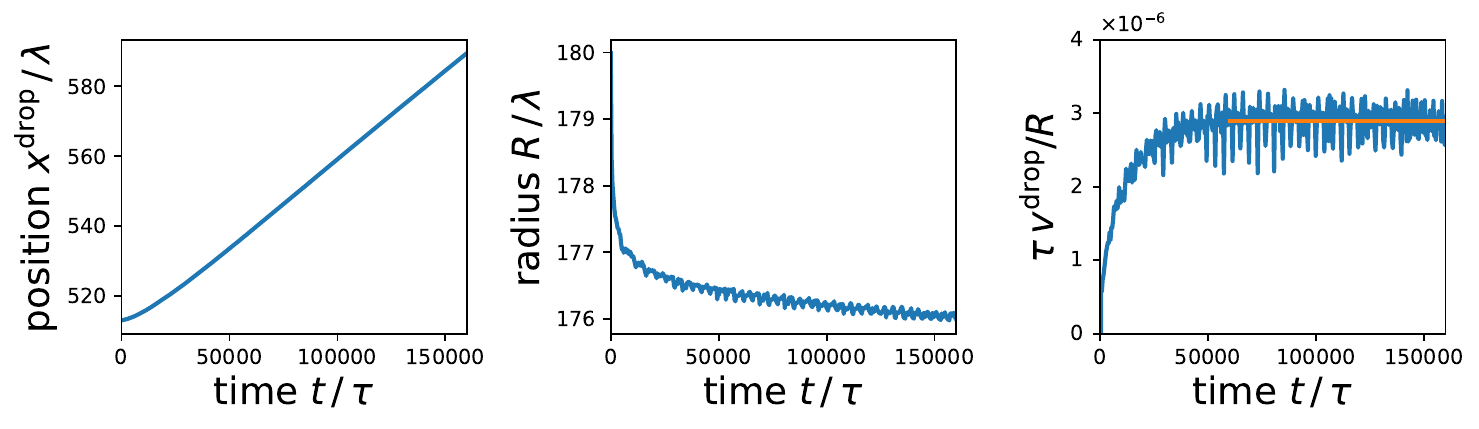}
    \caption{
        \textbf{Determination of normalized droplet velocity.}
        Droplet position (left panel), radius (middle panel), and normalized droplet velocity $v^\mathrm{drop} / R$ (right panel) as a function of time~$t$.
        Average normalized droplet velocity (orange line in right panel) is extracted by averaging normalized velocity over time after an initial transient.
        Model parameters are $R_0 = 180 \lambda$, $\Delta \bar{\mu}_R = 0.15 \, \kT$, $\bar\mu_R^+ = - 0.8 \, \kT$, $\chi_{DS} = 2.5$, $\chi_{DR} = 1.3$, $\chi_{RS} = 1.7$,
        $\tau = \lambda^2/(\Lambda_0 \ \kT)$, $\Lambda_{ij} = \Lambda_0 (\phi_i \delta_{ij} - \phi_i\phi_j)$,
        $L_x = 984 \, \lambda$, and $L_y = 1024 \, \lambda$.
    }
    \label{fig:Droplet_velocity_calc}
\end{figure}

To evaluate the droplet speed from numerical simulations, we first determine droplet positions~$\vect{x}^\mathrm{drop}_i$ (\Figref{fig:Droplet_velocity_calc}, left) and radii~$R_i$ (\Figref{fig:Droplet_velocity_calc}, middle) by fitting a spherical shape.
We do this using the \texttt{DropletTracker} class implemented in the \href{https://github.com/zwicker-group/py-droplets}{\textit{py-droplets}} Python package~\cite{pydropdoc}.
As a threshold for detecting the interface, we choose the mean of the minimum and maximum value of the droplet volume fraction $\phi_D$.
Additionally, the droplet position and radius are refined by fitting a diffuse spherical droplet~\cite{pydropdoc}.
We then determine the instantaneous velocity $\vect{v}^\mathrm{drop}_i = (\vect{x}^\mathrm{drop}_{i+1} - \vect{x}^\mathrm{drop}_{i})/(t_{i+1} - t_i)$ for various time points $t_i$ and average the normalized velocity $\vect{v}^\mathrm{drop}_i / R_i$ (\Figref{fig:Droplet_velocity_calc}, right) for a short time period after an initial transient (orange line in \Figref{fig:Droplet_velocity_calc}, right) to estimate the normalized droplet velocity $\vect{v}^\mathrm{drop}/R$.
Small fluctuations in the radius and velocity over time are due to lattice effects when fitting the spherical shape, but these do not affect the averaged velocity estimate.

\section{Estimation of chemical potential gradient inside the droplet}
\label{sec:estimation}

\begin{figure}
    \centering
    \includegraphics[width=0.45\linewidth]{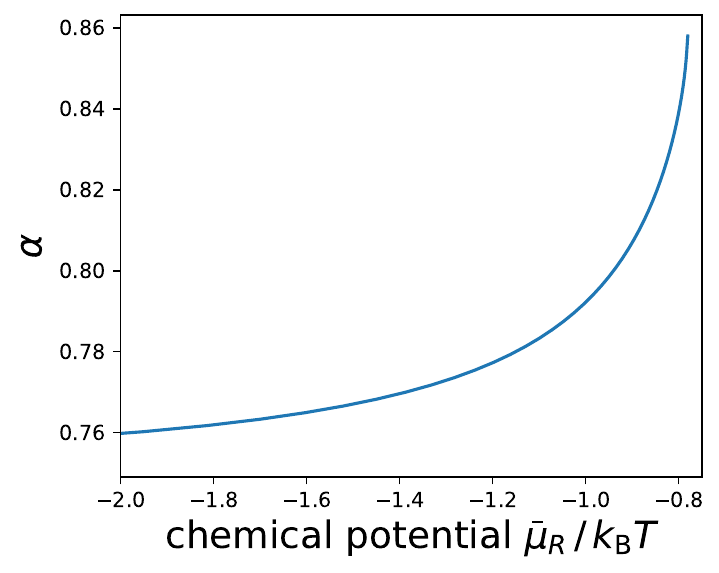}
    \caption{
        \textbf{Estimation of chemical potential gradient inside droplet.}
        Ratio~$\alpha$ of chemical potential gradient $\partial_x \bar\mu_R^\mathrm{in}$ inside the droplet and applied gradient $\Delta \bar\mu_R / L$ for different values of chemical potential $\bar\mu_R$ at the center of the droplet. Ratios are calculated with \Eqref{eq:ratio_gradients} using only equilibrium values extracted from the phase diagram.
        Model parameters are $\chi_{DS} = 2.5$, $\chi_{DR} = 1.3$, $\chi_{RS} = 1.7$, $\Lambda_{ij} = \Lambda_0 (\phi_i \delta_{ij} - \phi_i\phi_j)$.
    }
    \label{fig:alpha}
\end{figure}

Partitioning of regulator material into the respective phases alters the effective diffusion, and therefore, the local chemical potential gradient.
To estimate how partitioning affects the local gradient, we approximate the system by simple one-dimensional, quasi-stationary chemical potential profiles.
Linearizing the quasi-stationary diffusion equations, $\partial_x j_i = 0$, inside and outside the droplet, we get
\begin{subequations}
\begin{align}
    0 &= \Lambda_{DD,0}^\mathrm{in/out} \partial_x^2 \delta\bar{\mu}_D^\mathrm{in/out} + \Lambda_{DR,0}^\mathrm{in/out} \partial_x^2 \delta\bar{\mu}_R^\mathrm{in/out} \; , \\
    0 &= \Lambda_{RR,0}^\mathrm{in/out} \partial_x^2 \delta\bar{\mu}_R^\mathrm{in/out} + \Lambda_{DR,0}^\mathrm{in/out} \partial_x^2 \delta\bar{\mu}_D^\mathrm{in/out} \; .
\end{align}
\end{subequations}
Adding and subtracting the two equations, we can decouple them,
\begin{subequations}
\begin{align}
    0 &= \left(\Lambda_{DD,0}^\mathrm{in/out} - \frac{(\Lambda_{DR,0}^\mathrm{in/out})^2}{\Lambda_{RR,0}^\mathrm{in/out}} \right) \partial_x^2 \delta\bar{\mu}_D^\mathrm{in/out} \; , \\
    0 &= \left(\Lambda_{RR,0}^\mathrm{in/out} - \frac{(\Lambda_{DR,0}^\mathrm{in/out})^2}{\Lambda_{DD,0}^\mathrm{in/out}} \right) \partial_x^2 \delta\bar{\mu}_R^\mathrm{in/out} \; ,
\end{align}
\end{subequations}
with boundary conditions
\begin{subequations}
\begin{align}
    \delta\bar{\mu}_R^\mathrm{out}|_{x_\pm} &= \bar\mu_R^\pm - \bar\mu_R^0 \; ,\\
    \partial_x \delta\bar{\mu}_D^\mathrm{out}|_{x_\pm} &= - \frac{\Lambda_{DR,0}^\mathrm{out}}{\Lambda_{DD,0}^\mathrm{out}} \partial_x \delta\bar{\mu}_R^\mathrm{out}|_{x_\pm} \; .
\end{align}
\end{subequations}
Consequently, the chemical potential is a linear function of $x$ inside and outside the droplet.
Assuming that the total amount of material is conserved in the domain, the fluxes of regulator material flowing into/out of the system,
\begin{align}
j_R^\mathrm{out} = -\left(\Lambda_{RR,0}^\mathrm{out} - \frac{(\Lambda_{DR,0}^\mathrm{out})^2}{\Lambda_{DD,0}^\mathrm{out}} \right) \partial_x \delta\bar{\mu}_R^\mathrm{out} \; ,
\end{align}
must be equal to the flux inside the droplet,
\begin{align}
j_R^\mathrm{in} = - \Lambda_{RR,0}^\mathrm{in} \partial_x \delta\bar{\mu}_R^\mathrm{in} \; ,
\end{align}
where we use that the gradient in droplet exchange chemical potential does not change significantly within the droplet.
Additionally, from geometric arguments, we know
\begin{equation}
    \bar\mu_R^+ - \bar\mu_R^- = 2 R \ \partial_x \delta\bar\mu_R^\mathrm{in} + (L - 2 R) \ \partial_x \delta\bar\mu_R^\mathrm{out} \; ,
\end{equation}
where $L=x_+-x_-$ is the length of the system in the direction of the gradient.
Therefore, the chemical potential gradient inside the droplet reads
\begin{equation}
    \partial_x \delta\bar\mu_R^\mathrm{in} = \alpha \ \frac{\Delta\bar\mu_R}{L}
    \qquad \text{with} \qquad
     \alpha = \frac{L}{2 R + (L - 2 R) \frac{\Lambda_{RR,0}^\mathrm{in}}{\Lambda_{RR,0}^\mathrm{out} - (\Lambda_{DR,0}^\mathrm{out})^2 / \Lambda_{DD,0}^\mathrm{out}}} \; .
\end{equation}
This relation reveals that the internal regulator chemical potential gradient depends on the droplet radius~$R$, which implies a weak radius-dependence of the normalized velocity $v^\mathrm{drop} / R$.
To get an estimate independent of $R$, we take the limit of the system being much larger than the droplet ($L \gg R$) to arrive at
\begin{equation}
    \alpha \approx \frac{\Lambda_{RR,0}^\mathrm{out} - (\Lambda_{DR,0}^\mathrm{out})^2 / \Lambda_{DD,0}^\mathrm{out}}{\Lambda_{RR,0}^\mathrm{in}} \; .
    \label{eq:ratio_gradients}
\end{equation}
Using our specific mobility model, $\Lambda_{ij} = \Lambda_0 (\phi_i \delta_{ij} - \phi_i \phi_j)$, and the values of $\phi_D$ and $\phi_R$ in the dense and dilute phase extracted from the equilibrium phase diagram, we can determine the ratio $\alpha$ between the internal and the applied average gradient without detailed information from the simulations (\Figref{fig:alpha}).

\section{Equilibrium calculation of surface tension}
\label{sec:SIeqCalc}

\begin{figure}
    \centering
    \includegraphics[width=0.9\linewidth]{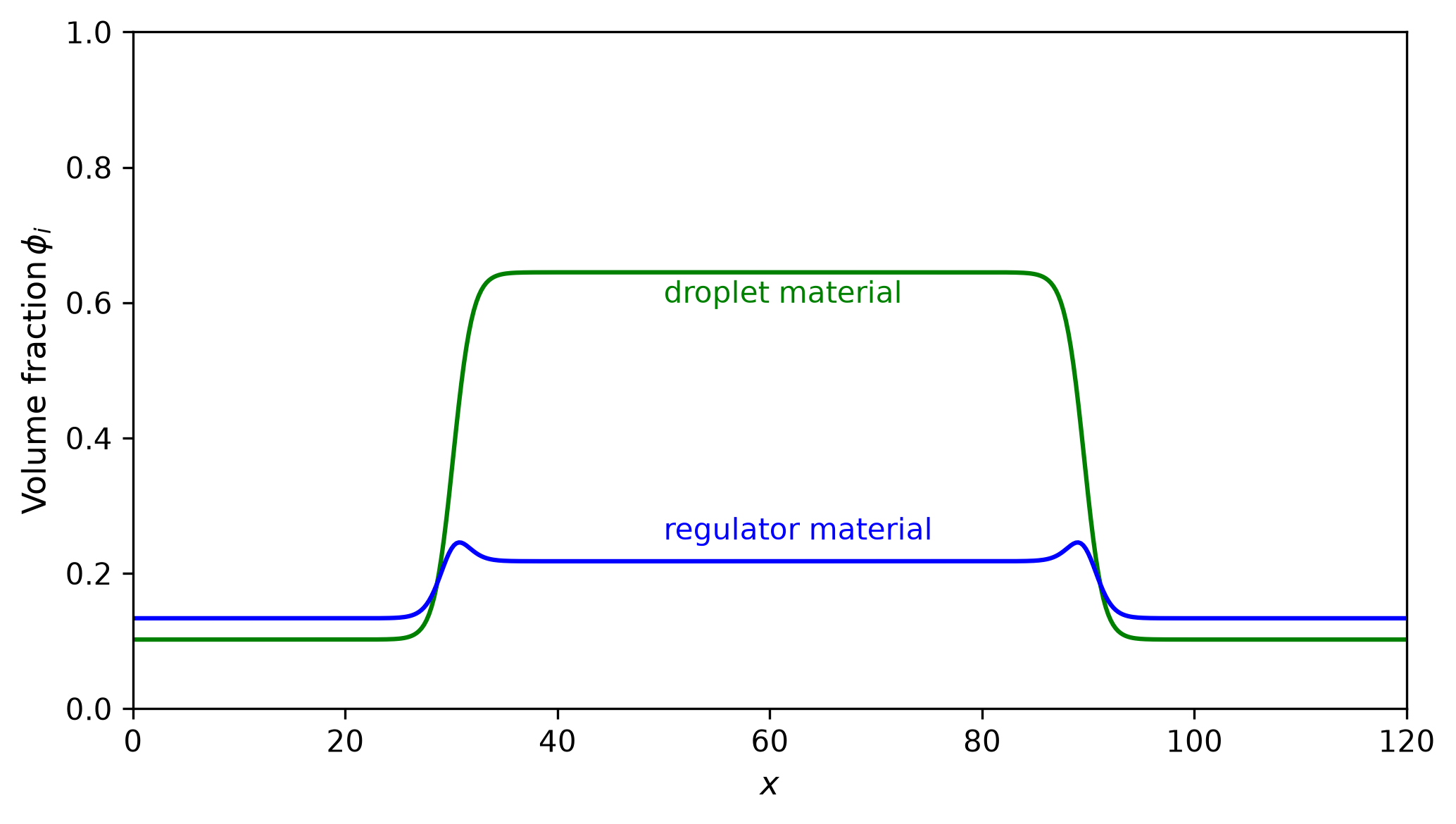}
    \caption{
        \textbf{Equilibrated profiles of droplet and regulator.}
        Representative equilibrated profiles of droplet and regulator volume fractions $\phi_D(x)$ and $\phi_R(x)$.
        Model parameters are $\chi_{DS} = 3$, $\chi_{DR} = 0.9$, $\chi_{RS} = 1.5$, and $\kappa_{ij} = -\lambda^2 (\chi_{ij} - \chi_{iS} - \chi_{Sj})$.
    }
    \label{fig:eqprofiles}
\end{figure}

We here describe how we obtain the surface tension associated with an equilibrium profile in the semi-grand-canonical setting.
For a prescribed $\bar\mu_R$, we first determine the two coexisting compositions based on the binodal conditions.
We use these concentrations to initialize a one-dimensional, periodic system with a  piecewise-linear profile containing two interfaces, which is compatible with periodic boundary conditions.
We then relax this system while conserving the resulting total amounts.
At equilibrium, the chemical potentials in both phases then match the prescribed semi-grand-canonical value.

To relax the profiles efficiently, we use globally conserved Model A dynamics,
\begin{equation}
\phi_i^{(n)}(t+\Delta t)
=
\phi_i^{(n)}(t)
-
\Delta t
\left(
\frac{\delta F}{\delta \phi_i^{(n)}}
-
\frac{1}{N_L}
\sum_{j=1}^{N_L}
\frac{\delta F}{\delta \phi_j^{(n)}}
\right)
\;,
\end{equation}
where $n=D, R$ labels the droplet and regulator species respectively, $i$ denotes the lattice site, and $N_L=512$ marks the total number of lattice sites.
Subtracting the spatially averaged functional derivative ensures conservation of the total amount of each species for all times,
\begin{equation}
\frac{1}{N_L}\sum_i \,\phi^{(n)}_i(t)=\bar{\phi}^{(n)}
\;.
\end{equation}
We use this constrained Model A relaxation instead of Model B dynamics because the latter introduces an additional Laplacian acting on the chemical potential and therefore higher-order spatial derivatives.
Since only the equilibrium state is required here, the constrained Model A dynamics provides a more efficient route in obtaining it.

Spatial derivatives are evaluated using the pseudospectral method. To improve numerical stability, each update is modified by a wavenumber dependent low-pass filter,
\begin{equation}
\Delta t
\rightarrow
\frac{\Delta t}{1+k^2\Delta t}
\label{eqn:spectral_filter}
\;,
\end{equation}
where $k$ is the wavenumber of the corresponding Fourier mode.
This filter suppresses high-wavenumber modes, which are typically responsible for the onset of numerical instabilities and allows larger time steps to be used.
The time step is system dependent and can become very small particularly close to critical points.
We evolve the system until
\begin{equation}
    \max_{i,n}
    \left|
        \phi_i^{(n)}(t+\Delta t)-\phi_i^{(n)}(t)
        \right|
        <10^{-10}
        \;,
\end{equation}
ensuring that stationary state is reached. 
Since the filter given by \Eqref{eqn:spectral_filter} only modifies the relaxation path, but not the fixed points, the equilibrium states are the same that one would have obtained for relaxation with fixed $\Delta t$.

Finally, the surface tension~$\gamma$ is  calculated from the equilibrated profiles according to \Eqref{eqn:surface_tension} in the main text.
Since the system contains two interfaces, we divide the integral by a factor of two, to obtain the surface tension of a single interface.
Representative equilibrated profiles are shown in \Figref{fig:eqprofiles}.

\section{Qualitative form of surface tension is largely independent of $\kappa_{ij}$}
\label{sec:SIkappa}

\begin{figure}
    \centering
    \includegraphics[width=\linewidth]{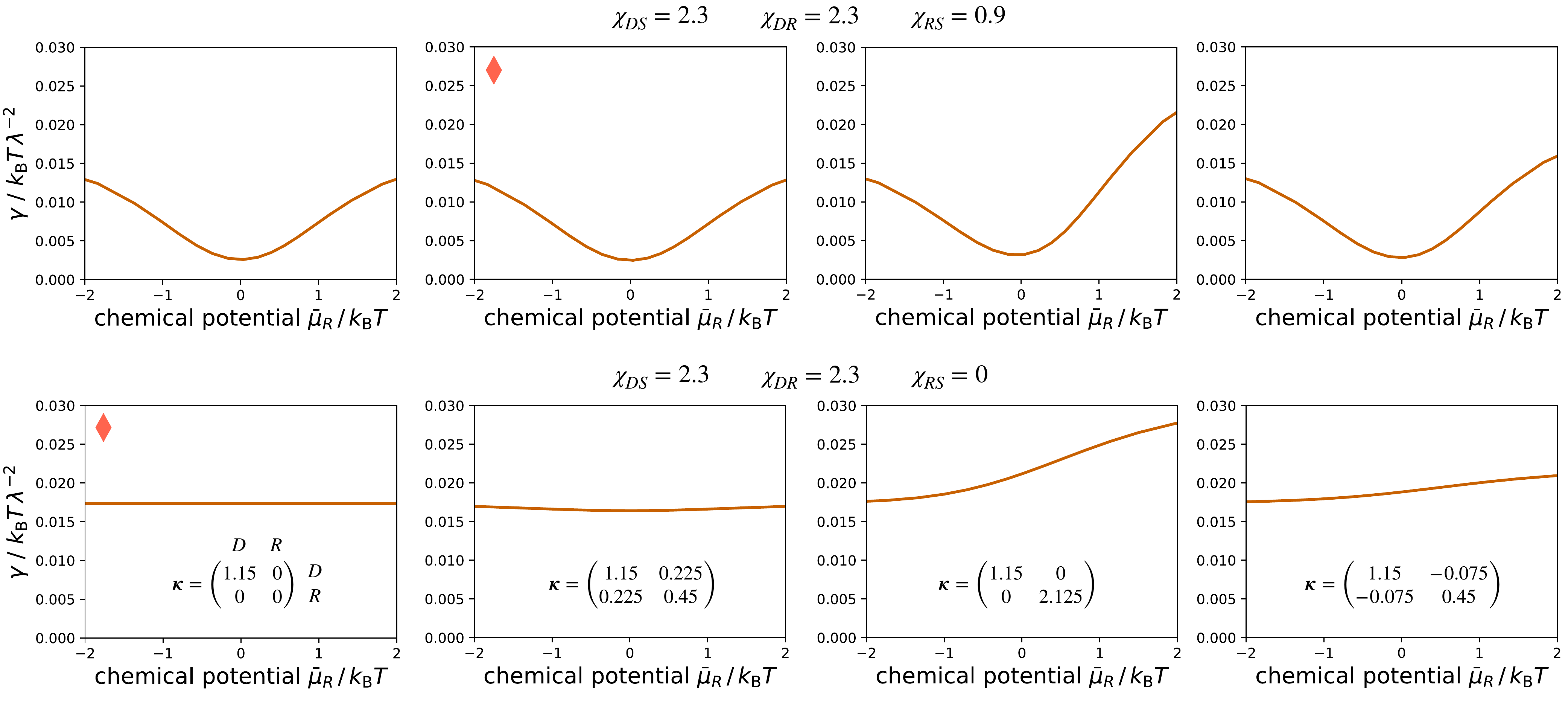}
    \caption{
        \textbf{Gradient matrix influences qualitative behavior only in the marginal case.}
        Surface tension~$\gamma$ as a function of chemical potential~$\bar\mu_R$ for the generic case without critical points (top row, \Figref{fig:phasediagrams}C in the main text) and the marginal case (bottom row, \Figref{fig:phasediagrams}D in the main text) for various $\kappa_{ij}$ (rows, values marked in lower  plots). %
        Note that the cases where $\kappa_{ij} = -\lambda^2(\chi_{ij} - \chi_{iS} - \chi_{Sj})$, that are presented in \Figref{fig:phasediagrams} (bottom row), are also shown (red diamonds).
        Model parameters are $\chi_{DS} = 2.3$, $\chi_{DR} = 2.3$, $\chi_{RS} = 0.9$ (top row), and $\chi_{DS} = 2.3$, $\chi_{DR} = 2.3$, $\chi_{RS} = 0$ (bottom row).
    }
    \label{fig:marginal}
\end{figure}

For phase diagram morphologies containing critical points (\Figref{fig:phasediagrams}A,B in main text), the qualitative form of the surface tension as a function of chemical potential is independent of the choice of $\kappa_{ij}$, because $\gamma$ must vanish at critical points and has to be positive otherwise.
While this shows that the phase diagram governs the qualitative behavior of $\gamma$, the precise choice of the interfacial penalty given by $\kappa_{ij}$ can affect $\gamma$ quantitatively.
Even in the absence of critical points (\Figref{fig:phasediagrams}C in main text), the qualitative form of the surface tension is generically determined by the morphology of the phase diagram as well, and is independent of $\kappa_{ij}$ (see \Figref{fig:marginal} top row).
However, for the marginal case (\Figref{fig:phasediagrams}D in main text), where we expect a constant surface tension, the choice of $\kappa_{ij}$ influences the qualitative behavior (see \Figref{fig:marginal} bottom row).
Consequently, we expect that the qualitative form of the surface tension is more sensitive to the choice of $\kappa_{ij}$ if the phase diagram is closer to the marginal case.

The conclusions drawn above assume a constant gradient matrix $\kappa_{ij}$, which is the case for the model considered in this work. 
For complex molecules, $\kappa_{ij}$ can be more complicated and may depend on the local composition $\phi_i$, which may lead to more complex behavior.
However, we expect that the qualitative behavior of the surface tension is still largely determined by the morphology of the phase diagram, specifically by critical points, where $\gamma$ must vanish independent of the choice of $\kappa_{ij}$.

\end{document}